\documentclass[conference]{IEEEtran}
\IEEEoverridecommandlockouts
\usepackage{cite}
\usepackage{amsmath,amssymb,amsfonts}
\usepackage{algorithmic}
\usepackage{graphicx}
\usepackage{textcomp}
\usepackage[dvipsnames]{xcolor}
\usepackage[colorinlistoftodos,prependcaption]{todonotes}
\usepackage[utf8]{inputenc}  \usepackage{graphicx}
\usepackage[numbers]{natbib}
\usepackage[T1]{fontenc}    
\usepackage{hyperref}       
\usepackage{url}            
\usepackage{booktabs}       
\usepackage{amsfonts}       
\usepackage{nicefrac}       
\usepackage{microtype}      
\usepackage{pifont}
\usepackage{amssymb}
\usepackage{caption}
\usepackage{xspace}
\usepackage{tabularx}
\usepackage{subfig}

\newcolumntype{b}{X}
\newcolumntype{s}{>{\hsize=.5\hsize}X}

\newcommand{\alg}{\textsc{DFA}\xspace}
\newcommand{\algl}{\textsc{DFA-R}\xspace}
\newcommand{\algg}{\textsc{DFA-G}\xspace}
\newcommand{\algd}{\textsc{RefD}\xspace}
\newcommand{\asr}{\textit{ASR}\xspace} 
\newcommand{\dpr}{\textit{DPR}\xspace} 
\newcommand{\cmark}{\ding{51}}%
\newcommand{\xmark}{\ding{55}}%
\def\BibTeX{{\rm B\kern-.05em{\sc i\kern-.025em b}\kern-.08em
    T\kern-.1667em\lower.7ex\hbox{E}\kern-.125emX}}

\newcommand{\changesr}[1]{\textcolor{black}{#1}}

\begin{document}

\title{Fabricated Flips: \\ Poisoning Federated Learning without Data\thanks{Published in DSN 2023.}}

\author{\IEEEauthorblockN{Jiyue Huang, Zilong Zhao, Lydia Y. Chen, Stefanie Roos}
\IEEEauthorblockA{TU Delft, The Netherlands. \{J.Huang-4, Z.Zhao-8, Y.Chen-10, S.Roos\}@tudelft.nl} 
}

\maketitle

\begin{abstract}

Attacks on Federated Learning (FL) can severely reduce the quality of the generated models and limit the usefulness of this emerging learning paradigm that enables on-premise decentralized learning. 
However, existing untargeted attacks are not practical for many scenarios as they assume that \textit{i)} the attacker knows every update of benign clients, or \textit{ii)} the attacker has a large dataset to locally train updates imitating benign parties. 

In this paper, we propose a data-free  
untargeted attack (\alg) that synthesizes malicious data to craft adversarial models without eavesdropping on the transmission of benign clients at all or requiring a large quantity of task-specific training data. 
We design two variants of \alg, namely 
\algl and \algg, which differ in how they trade off stealthiness and effectiveness. 
Specifically,  \algl iteratively 
optimizes a malicious data layer to minimize the prediction confidence of all outputs of the global model, whereas \algg interactively trains a malicious data generator network by steering the output of the global model toward a particular class. 
Experimental results on Fashion-MNIST, Cifar-10, and SVHN show that \alg, despite requiring fewer assumptions 
than existing attacks, achieves similar or even higher attack success rate than state-of-the-art untargeted attacks against various state-of-the-art defense mechanisms. Concretely, they can evade all considered defense mechanisms in at least 50\% of the cases for CIFAR-10 and often reduce the accuracy by more than a factor of 2. 

Consequently, we design \algd, a defense specifically crafted to protect against data-free attacks. 
\algd leverages a reference dataset to detect updates that are biased or have a low confidence. 
It greatly improves upon existing defenses by filtering out the malicious updates and achieves high global model accuracy. 

\end{abstract}


\section{Introduction}



Federated learning (FL)~\cite{DBLP:journals/tist/YangLCT19,DBLP:conf/icdcs/NguyenMMFAS19} enables distributed training of machine learning models, e.g., multi-class image classifiers, without sharing the raw data. \textit{Clients} train models locally and the overall model, called the global model, is an aggregation of these local models.
The training proceeds in multiple rounds: in each round, \textit{the central server} provides a global model that clients use to initialize their local models. They then train on their local dataset and provide updates to the central server, who aggregates these updates to a new global model for the next round. 
In this manner, models requiring personal data such as information about medical or financial conditions can be obtained without explicit privacy violations. 
Recently, FL has been applied to domains such as detection of credit card fraud~\cite{fraud:conf/ijcai/ZhengYG020,credit:journals/corr/abs-2108-07927}, cybersecurity center operations~\cite{cyber:conf/icdcs/KhramtsovaHLS20}, and medical relation extraction~\cite{medical:conf/emnlp/SuiCZJXS20}. 

A downside of preventing the central server from accessing local data is that it limits the ability to detect misbehavior. Adversarial clients may reduce the quality of the model by manipulating the  data they train on~\cite{labelflip:conf/esorics/TolpeginTGL20} or their local model directly~\cite{Fang:conf/uss/FangCJG20}.
\changesr{Cross-device~\cite{cross-device:karimireddy2021breaking,hard2018federated} FL, which allows arbitrary 
parties to join the distributed training, is especially vulnerable as attackers can easily infiltrate the system.}
The attack can be untargeted, i.e., aiming for an overall accuracy degradation of the trained global model.
It can also be targeted, i.e., only supposed to affect certain input, e.g., inject backdoors that lead to wrong model output from input data with a certain chosen feature~\cite{backdoor:conf/aistats/BagdasaryanVHES20}. \changesr{In this paper, we focus on untargeted attacks, as they are far-reaching denial-of-service attacks. In cross-device FL, attackers may run such a denial-of-service attack to undermine a competing company from getting meaningful models after their users. Furthermore, when machine learning as a service~\cite{ribeiro2015mlaas} is extended to include FL~\cite{kourtellis2020flaas}, untargeted attacks aiming to cause losses for a service provider are to be expected, similar to current denial-of-service attacks on  Amazon Web Services and Github~\footnote{\tiny{\url{https://www.a10networks.com/blog/aws-hit-by-largest-reported-ddos-attack-of-2-3-tbps/}}}}.

There have been a number of untargeted attacks on FL~\cite{Fang:conf/uss/FangCJG20,LIE:conf/nips/BaruchBG19,ndss:conf/ndss/ShejwalkarH21}. 
Yet, some attacks~\cite{LIE:conf/nips/BaruchBG19,ndss:conf/ndss/ShejwalkarH21} assume that the adversary is aware of all of the updates that benign clients send. It is unclear how they can practically obtain such knowledge as clients only share the updates with the benign central server and communication can be encrypted to prevent eavesdropping from the adversary.  
While not all attacks require benign updates, attacks that can succeed without this knowledge requires that the attacker has a considerable amount of training data to train substitute benign updates~\cite{Fang:conf/uss/FangCJG20}.
Although this assumption is realistic for common tasks, e.g., image classification of common pets, the possession of such data is much less likely for special-purpose tasks, e.g., classification of rare disease based on detailed medical data~\cite{rare:journals/corr/abs-2003-08119}.

In this paper, we consider whether it is actually necessary to have real data (or benign updates). \changesr{One may expect that in cross-device FL,
it is relatively easy to obtain data as everyone can join, which might indicate that everyone can have data. However, there exist scenarios where admission is not restricted to a predefined group because there are few parties that can contribute and it is not known who they are. For instance, for a study on the live of people with a rare disease, it might not be possible to access medical records on who has the disease, so it makes sense to just publicly ask for participation. Furthermore, not requiring parties to identify before joining allows them to participate anonymously, possibly using tools like Tor~\cite{dingledine2004tor} to send in their updates without having to fear that they reveal that they have a certain medical condition, which could increase their insurance premium or prevent them from gaining employment.  In such a scenario, it is also hard to corrupt participating clients and use their data, as the identity of the clients is not known. 
Even if the learning task is such that is easy to obtain data, e.g., a software company aiming to build a model on how users interact with their tool, it is still additional overhead for the attackers, e.g., they have to either use the tool themselves or obtain data from a real user. Thus, even if the attacker can get data, the question of whether they \emph{have to} or can skip the overhead of data acquisition is essential as without data acquisition, it is more likely that attacks can be automated and run at scale against many FL learning tasks. }

We design 
a novel Data-Free Attack (\alg) and evaluate it on the example of image classification. The goal of the attack is to reduce the overall accuracy of the model through the injection of malicious model updates based on synthetic images. 
In each round, the attacker first generates malicious images by making use of the received global model and then trains the local adversarial model using those images paired with a randomly chosen class $\tilde{Y}$. 
We design two variants of \alg, \algl and \algg, which steer the global model to classify images to either have low confidence or to classify incorrectly.  
Our first attack variant, \algl, generates synthetic local data by adding a {filte\textbf{R}} layer to the training. This data generation optimizes towards local synthetic data that is ambiguous according to the current global model, i.e., the current global model should output each of the $L$ possible classes with equal probability. 
A local model corresponding to such data diverts the global model and reduces classification accuracy. 
In contrast, our second attack, 
\algg, iteratively trains a {\textbf{G}enerator} that should produce synthetic images that are not from a specific randomly chosen class $\tilde{Y}$. We then assign these images with class label $\tilde{Y}$ and train on the resulting dataset, thus implicitly combining synthetic data generation with label flipping for poisoning. 

To improve stealthiness for both attacks, we add a regularization term to the loss function of the classifier that steers the update generation such that updates are not detected as outliers and hence not removed by defenses. \alg thus stealthily bypasses the defense by ensuring that the deviation to the global model follows similar patterns as benign updates. 

In our evaluation, we determine the attack success rate, i.e., the decrease in model accuracy caused by the attack, and the rate at which our attackers pass the defense. 
We evaluate different levels of data heterogeneity by assigning data to clients according to the 
Dirichlet distribution,  which is a common model for heterogeneous real-world distributions~\cite{dirichlet:journals/bioinformatics/WuSWCC17}.
\algl and \algg reduce the accuracy of the trained model by a factor of 2 for most settings, even if defenses are applied.  
In comparison to state-of-the-art attacks, \algl and \algg achieve similar results, despite having weaker assumptions. Indeed, for most scenarios, our attacks perform slightly better than the existing attacks. 


Having shown that data-free attacks have severe impact on the accuracy of FL, we 
propose a defense strategy, \algd, which aims to defend against \algg and \algl by leveraging a reference dataset at the server. 
Based on this reference dataset, the central server determines whether a received model update is biased toward a certain class, which is typical for \algg, or shows a low confidence, which is typical for \algl. It combines these two factors into a novel defense score, termed $\mathcal{D}$-score.
Our evaluation results show that \algd successfully defends against the proposed data-free attacks, achieving accuracies that are close to the accuracy achieved in the absence of both attacks and defenses.




\section{Background and Related Work}
\begin{table*}[t!]
    \centering
    \begin{tabular}{c|c|c|c|c|c}
    \toprule
    Attacks & LIE\cite{LIE:conf/nips/BaruchBG19} &  Fang\cite{Fang:conf/uss/FangCJG20} &  Min-Max\cite{ndss:conf/ndss/ShejwalkarH21} & Min-Sum\cite{ndss:conf/ndss/ShejwalkarH21}& \alg (ours)\\
    \midrule
    No benign updates needed  & \xmark &  \cmark/\xmark  & \cmark/\xmark & \cmark/\xmark  & \cmark \\
    Defense-agnostic &\cmark & \xmark & \cmark & \cmark & \cmark\\
    No raw data needed & \cmark & \cmark/\xmark  & \cmark/\xmark & \cmark/\xmark & \cmark \\
    Heterogeneity considered & \xmark & \cmark & \cmark & \cmark & \cmark\\
    Attack type & Statistic & Statistic & Statistic & Statistic &Optiminzation\\
  \bottomrule
    \end{tabular}
    \caption{Attack scenarios in the state-of-the-art and ours.}
    \label{tab:knowledge}
    \vspace{-2em}
\end{table*}
\subsection{Federated Learning Primer}
As a distributed machine learning framework, FL systems consist of a set of $N$ clients and a central server. The global training process considers $R$ consecutive rounds indexed by the round number $t$. After model initialization by the server, each client $i$ (for $i = 1, 2,..., N$) trains a local model based on their own real data without sharing the raw data. The server iteratively aggregates models/gradients submitted from clients and distributes the aggregated model to the clients until reaching global model convergence. 
As clients can be offline or unresponsive, only a subset of them usually submits updates. 

In this paper, we focus on image classification tasks with $L$ classes.
Let $D_i$ be the local dataset of client $i$ and $F$ be the objective function for the classification task. The client $i$ updates its local model weights based on the global model $\boldsymbol{w}(t)$ by:
\begin{equation}
    \boldsymbol{w}_i(t+1) = \boldsymbol{w}(t)-\eta\frac{\partial{F(\boldsymbol{w}(t),D_i)}}{\partial{\boldsymbol{w}(t)}},
\end{equation}
where $\eta$ is the global uniformed learning rate.

For aggregating models of $K \leq N$ clients, the predominant method for attack-free scenarios is FedAvg~\cite{fedavg:conf/aistats/McMahanMRHA17}, which aggregates the new global model as a weighted average of the submitted local models \changesr{, i.e., 
\begin{equation}\label{eq:aggregation}
    \boldsymbol{w}(t)=\sum_{i=1}^{K} \frac{n_{i}}{\sum_{k=1}^{K} n_{k}} \boldsymbol{w}_{i}(t),
\end{equation}
where $n_i$ is the number of training samples of client $i$. }
However, the above algorithm is not robust under attacks~\cite{backdoor:conf/aistats/BagdasaryanVHES20,Fang:conf/uss/FangCJG20,ndss:conf/ndss/ShejwalkarH21,DBA:conf/iclr/XieHCL20}, hence defenses for securing the aggregation against maliciously crafted updates (also called robust aggregation methods) have been developed. 

\subsection{Existing attacks in FL systems}

FL empowers clients by leaving the training to them and not revealing the local data. However, as a consequence, FL systems are vulnerable to malicious behaviors.
Attacks can happen during the \textbf{training time}~\cite{LIE:conf/nips/BaruchBG19,ndss:conf/ndss/ShejwalkarH21,backdoor:conf/aistats/BagdasaryanVHES20,Fang:conf/uss/FangCJG20,DBA:conf/iclr/XieHCL20} or \textbf{inference time}~\cite{reconstruction:conf/cvpr/YinMVAKM21,userlevel:journals/jsac/SongWZSWRQ20,membership:conf/sp/NasrSH19}.
For the inference-time attacks, attackers aim to infer private data~\cite{membership:conf/sp/NasrSH19}. They may even reconstruct the private local training data~\cite{reconstruction:conf/cvpr/YinMVAKM21}.
In this paper, we focus on training-time attacks where attackers participate in the training. We classify the training-time attacks from two perspectives: \textit{i}) the attack objectives and \textit{ii)} the attacked component of the FL system, e.g., data or model.

There are three attack objectives for training-time attacks:  \textbf{Free-riding}~\cite{DBLP:journals/corr/freeriderlin,fraboni2021free} is used to obtain the global model without contributing data and computation.
\textbf{Targeted attacks}~\cite{backdoor:conf/aistats/BagdasaryanVHES20,DBA:conf/iclr/XieHCL20} aim to decrease the model accuracy for specific data, e.g., data with designed triggers. 
\textbf{Untargeted attacks}~\cite{LIE:conf/nips/BaruchBG19,Fang:conf/uss/FangCJG20,ndss:conf/ndss/ShejwalkarH21}, in contrast, aim to decrease the general accuracy of the model. 

There are four state-of-the-art untargeted attacks, namely LIE~\cite{LIE:conf/nips/BaruchBG19}, Fang~\cite{Fang:conf/uss/FangCJG20} as well as Min-Max and Min-Sum~\cite{ndss:conf/ndss/ShejwalkarH21}, which are two variants of the same attack idea. We summarize their key differences in Table~\ref{tab:knowledge}. 
All attacks require knowledge of the models of benign clients, real data, or knowledge of any defenses applied by the server. Some of them, like Min-Max, are flexible in that they can work with either benign updates or real data but they need at least one of the two, which we indicate by \cmark/\xmark{}  in the respective rows in the table. 
In terms of methods, all existing attacks rely on statistical methods or heuristics to construct the malicious updates by shifting the mean of benign updates without being detected. Concretely, LIE~\cite{LIE:conf/nips/BaruchBG19} calculates the mean and standard deviation of all of the benign updates and
then shifts the true mean by changing the value in one direction in such a manner that it is within the range that is considered acceptable by the defense. 
Shejwalkar et. al~\cite{ndss:conf/ndss/ShejwalkarH21} 
further improve LIE by adapting the scaling factor $z$ of the weighted sum as well as extending the standard deviation to the sign and unit vector of the gradient. By such means, the maximum distance (or the sum of squared distances for Min-Sum)  of the malicious gradient from all the benign gradients is upper bounded.
Note that while the authors~\cite{ndss:conf/ndss/ShejwalkarH21} propose a number of attacks according to different levels of adversarial knowledge, we only compare to the Min-Max attack, which is the strongest in their paper. 
Fang et. al~\cite{Fang:conf/uss/FangCJG20} propose an attack that
steers global model parameters in the opposite direction of the benign updates and ensures its stealthiness through the knowledge of  the exact defense. Aforementioned untargeted attacks, except LIE, are evaluated in junction with the heterogeneous data, which is 
attribute skewed~\cite{emnist:journals/corr/CohenATS17,multiview:conf/iccv/SuMKL15} or label skewed~\cite{noniid:conf/iclr/WangYSPK20,noniid:conf/icml/YurochkinAGGHK19,noniid:conf/nips/LinKSJ20}.

Moreover, attacks can also be categorized by the component which attacks act upon: data or model. During the training time, the adversary may inject malicious data with dirty labels or data to train the local model, e.g., label flipping~\cite{labelflip:conf/esorics/TolpeginTGL20} and trigger injection~\cite{backdoor:conf/aistats/BagdasaryanVHES20,DBA:conf/iclr/XieHCL20}. For example, backdooring\cite{backdoor:conf/aistats/BagdasaryanVHES20} is executed by injecting trigger-based malicious samples~\cite{backdoor:conf/aistats/BagdasaryanVHES20,Curse:conf/aaai/ZawadAC00BT021} into the local training dataset. DBA~\cite{DBA:conf/iclr/XieHCL20} then extends the study~\cite{backdoor:conf/aistats/BagdasaryanVHES20} to bypass Sybil defenses such as FoolsGold~\cite{FoolsGold:conf/raid/FungYB20}. 
Modeling poisoning~\cite{LIE:conf/nips/BaruchBG19,ndss:conf/ndss/ShejwalkarH21,Fang:conf/uss/FangCJG20,backdoor:conf/aistats/BagdasaryanVHES20,DBA:conf/iclr/XieHCL20} manipulates the submitted model rather than merely adopting malicious data to train, e.g., submit updates of the reversed sign of training gradient~\cite{Fang:conf/uss/FangCJG20}. Generally, model poisoning attacks require sophisticated technical capabilities such as eavesdropping and sufficient computation resources.


None of the existing attacks can deal with an attacker that does not have data unless they can observe the communication in plaintext. 
\changesr{Our attack uses a generator, as do other attacks but for highly different scenarios or goals: attacking centralized learning~\cite{mlattack:conf/cvpr/00810X0DZ022,mlattack:conf/iccv/ZhangBKK21}, attacking privacy~\cite{ganattack:conf/icc/ZhangZCY20, ganattack:journals/jsac/SongWZSWRQ20}, or mimicking prototypical samples of the other participants' training set with the goal of targeted attacks~\cite{ganattack:journals/iotj/ZhangCCBY21, ganattack:conf/trustcom/ZhangCWCY19}. }






\subsection{Existing Defense Mechanism in FL}
\label{sec:defensesSota}
\changesr{We here focus on defense mechanisms for FL, as algorithms designed for centralized learning (e.g., ~\cite{depois:journals/tifs/ChenZZWL21,Blacklight:conf/uss/LiSWZ0Z22} are not directly applicable in FL.} To tackle the attacks on FL systems, existing defense strategies can be conducted either on the server-side~\cite{krum:conf/nips/BlanchardMGS17,bulyan:conf/icml/MhamdiGR18,median:conf/icml/YinCRB18}, or the client-side~\cite{clientside:conf/nips/SunLDHCL21,clientside:journals/tdsc/ZhaoHWJSLH21,clientside:journals/corr/abs-2009-03561}. Server-side defenses are effective against both targeted and untargeted attacks due to the access to all model updates, whereas the state-of-the-art client-side defenses are merely shown to be effective against targeted attacks.
As we are concerned with untargeted attacks, we hence focus on server-side defenses.

Generally, there are three categories of server-side defenses: \textit{i) Sybil defenses} aim at detecting Sybil attackers who are controlled by one entity and submit similar updates. For example, \textit{FoolsGold}~\cite{FoolsGold:conf/raid/FungYB20} identifies Sybils based on the diversity of client contributions using cosine similarity of client updates. \textit{ii) Statistic defenses} curate the aggregated model by computing the statistics of every parameter across multiple updates. \textit{Median}~\cite{median:conf/icml/YinCRB18} utilizes the median value of all updates for each parameter whereas Trimmed mean (\textit{TRmean})~\cite{median:conf/icml/YinCRB18} excludes the minimum and maximum value from the average of each parameter.  \textit{iii) Outlier  detection}~\cite{krum:conf/nips/BlanchardMGS17,bulyan:conf/icml/MhamdiGR18} removes updates based on the pairwise distances of returned models. Higher distance implies that data owned by a client is of low quality or unrelated to the training task. \textit{Krum}~\cite{krum:conf/nips/BlanchardMGS17} only uses one update sent from the client whose cumulative distance of updates to the other updates is the lowest, taking the squared L2 Norm as a metric.  \textit{mKrum}~\cite{krum:conf/nips/BlanchardMGS17} extends this idea by choosing multiple updates.
\textit{Bulyan}~\cite{bulyan:conf/icml/MhamdiGR18} first selects updates
using \textit{mKrum} and further computes the trimmed mean of the selected gradients. 

\section{An Optimization-based Data-free Attack}
\label{sec:dfa}


\begin{figure}[ht!]
\centering
\includegraphics[width=0.45\textwidth]{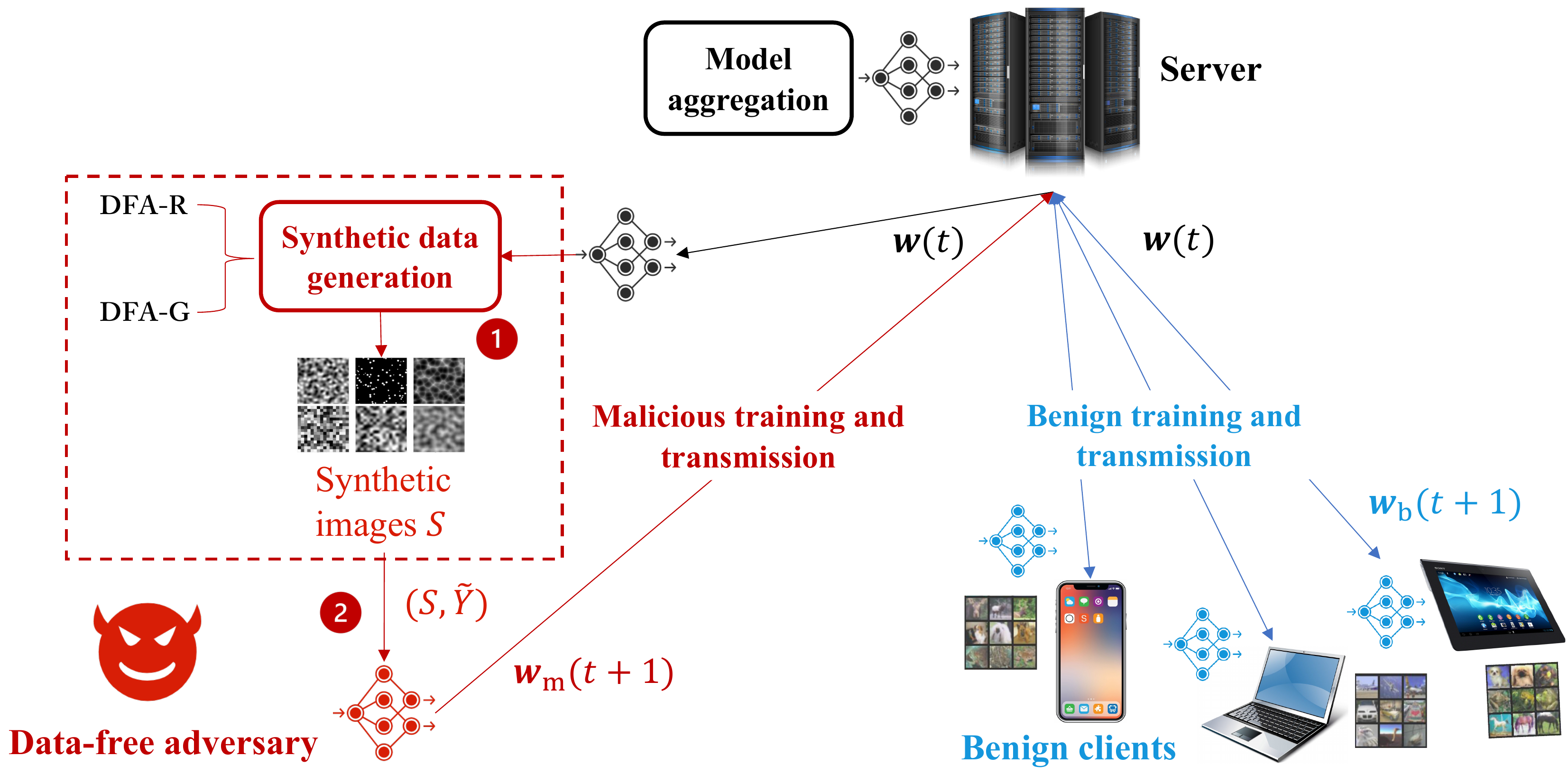}
\caption{
The framework of our proposed data-free attack (\alg) without knowing benign updates and owning raw data.}
\label{img:framework}
\end{figure}

In this section, we first introduce the threat model for our work. Then we propose our data-free attack (\alg) with two variants to generate malicious data inputs and local model updates: \algl and \algg. 

\subsection{Threat Model}
 \label{sec:dfa_threat_model}
We assume that communication between clients and the central server uses 
encrypted and authenticated channels, which prevent eavesdropping and manipulation of data during transmission. As a consequence, attackers are unaware of benign client updates. 
Benign clients always follow the protocol whereas malicious clients may arbitrarily deviate. All attackers may submit the same update. We add these assumptions for simplicity as we can easily circumvent Sybil defenses by adding small perturbation noise, as shown in the related work~\cite{backdoor:conf/aistats/BagdasaryanVHES20}.
The central server applies a defense mechanism, which is not known to the clients.

\changesr{We focus on cross-device FL, which means that anyone can join and at the same time there is client selection each round. Furthermore, the adversary inserts their own clients in the system rather than corrupt other clients. Corrupting other clients requires knowing the identities of other clients, which is not explicitly shared in cross-device FL. In the absence of anonymous communication, the adversary could obtain the identities only from observing network traffic but the ability to observe network traffic is restricted to internet service providers and other parties, so we do think it is more realistic to assume that the adversary does not know the other clients and hence also cannot easily corrupt them.}



Additionally, we assume that malicious parties do not 
have any data so as to enhance the versatility of the adversary. In practice, the difficulty of obtaining data varies between tasks. It is reasonable to assume that there are tasks relying on rare data that an attacker cannot easily obtain. 
We assume that all computations are executed by one adversarial party, who then sends the updates to individual malicious clients.

\textbf{Objectives: }The overall objective of an untargeted attack 
in Federated Learning is reducing the accuracy of the global model maintained by the central server. As a part of achieving this objective, clients need to craft malicious updates that bypass the applied defense. 


\textbf{Capabilities: }First, we assume that the number of malicious users controlled by the adversary in the system
does not exceed 50\% of the total clients. It seems implausible that a defense can overcome a higher number of attackers as defenses typically need a reference for benign behavior. 
The attacker cannot break cryptographic primitives. More generally, it is computationally bounded so that it cannot solve NP-complete or NP-hard problems. Otherwise, they can arbitrarily control the communication and computation of the malicious clients but not of any other parties in the system.  

\textbf{Knowledge: }Neither the defense algorithm nor benign updates are known to
the adversary. 
As the attacker also does not have data, the only knowledge of the adversary is the classification task in general, i.e., the number of classes, which is necessarily accessible as the server distributes the model.  

\subsection{Attack Optimization Framework }
The overall framework of our proposed attack \alg is illustrated in Fig.~\ref{img:framework}.
The server first distributes the current global model (classifier) $\boldsymbol{w}(t)$ to all of the clients. The benign clients truthfully follow the protocol and send the trained model $\boldsymbol{w}_b(t+1)$ back to the server. Malicious clients send the adversarial model $\boldsymbol{w}_m(t+1)$ instead. Then the server aggregates the submitted updates according to the deployed defense. As attackers do not have real data or benign updates, intuitively, the most obvious approach to attack is to directly change $\boldsymbol{w}(t)$. 


We experimented with using random weights but the attack was detected almost always. Concretely, only 2.62\% and 6.57\% of all updates submitted by malicious clients with random model weights bypassed the \textit{mKrum} defense for Fashion-MNIST and Cifar-10, respectively. For the \textit{Bulyan} defense, the attack only bypassed the defense in 3.27\% of the cases for Fashion-MNIST and always failed for Cifar-10. 
As manipulating the model directly does not seem a promising approach, 
we optimize the generation of synthetic malicious images according to $\boldsymbol{w}(t)$ and then use it to train the local adversarial model every round. The attack process consists of the following two steps.

\textbf{1. Malicious image generation.} 
We propose two optimization methods to synthesize malicious images based on different optimization methods, objectives of adversarial models and, importantly, the feedback of the global model.
The first method is \textbf{\algl}, which introduces an additional \textbf{filter input layer}\footnote{Such a layer has the same input dimension as the original image.} and optimizes it from a dummy image with the objective to  reduce the confidence on all  outputs of the global model. 
Our second method, \textbf{\algg}, designs a \textbf{generator network} to synthesize malicious images such that the output of the global model biases toward a randomly chosen class. 
As such, the generated noisy images paired with incorrect labels are applied to malevolently update the current model. 
The details on \algl and \algg are discussed in the following subsections. 

\textbf{2. Adversarial classifier training with distance-based loss.} In this step, the attacker uses synthetic data as generated by step 1
to train the classifier $\boldsymbol{w}_m(t+1)$. The optimization problem of the attack then becomes $\min_{\boldsymbol{w}_i}F(\boldsymbol{w}_i, S)$, where $S$ 
is the generated image set.
In order to enhance the stealthiness and hence pass the unknown defense, we propose to train with a distance-based loss function $\min_{\boldsymbol{w}_i} (F(\boldsymbol{w}_i, S)+\mathcal{L}_d)$ with regularization term $\mathcal{L}_d$ to enhance stealthiness
(detailed illustration in Sec.~\ref{subsec:distance_loss}). 
The size of $S$, $|S|$, is a hyper parameter of our attack framework that depends on the task. In the evaluation, we find that using a similar number of images as benign clients results in an effective attack.
The adversary can estimate the size during training based on the aggregated results of the global model and the duration that other clients require for training. 

\vspace{-0.5em}
\subsection{\algl synthetic data generation}

\begin{figure}[htpb!]
\centering
\includegraphics[width=0.75\columnwidth]{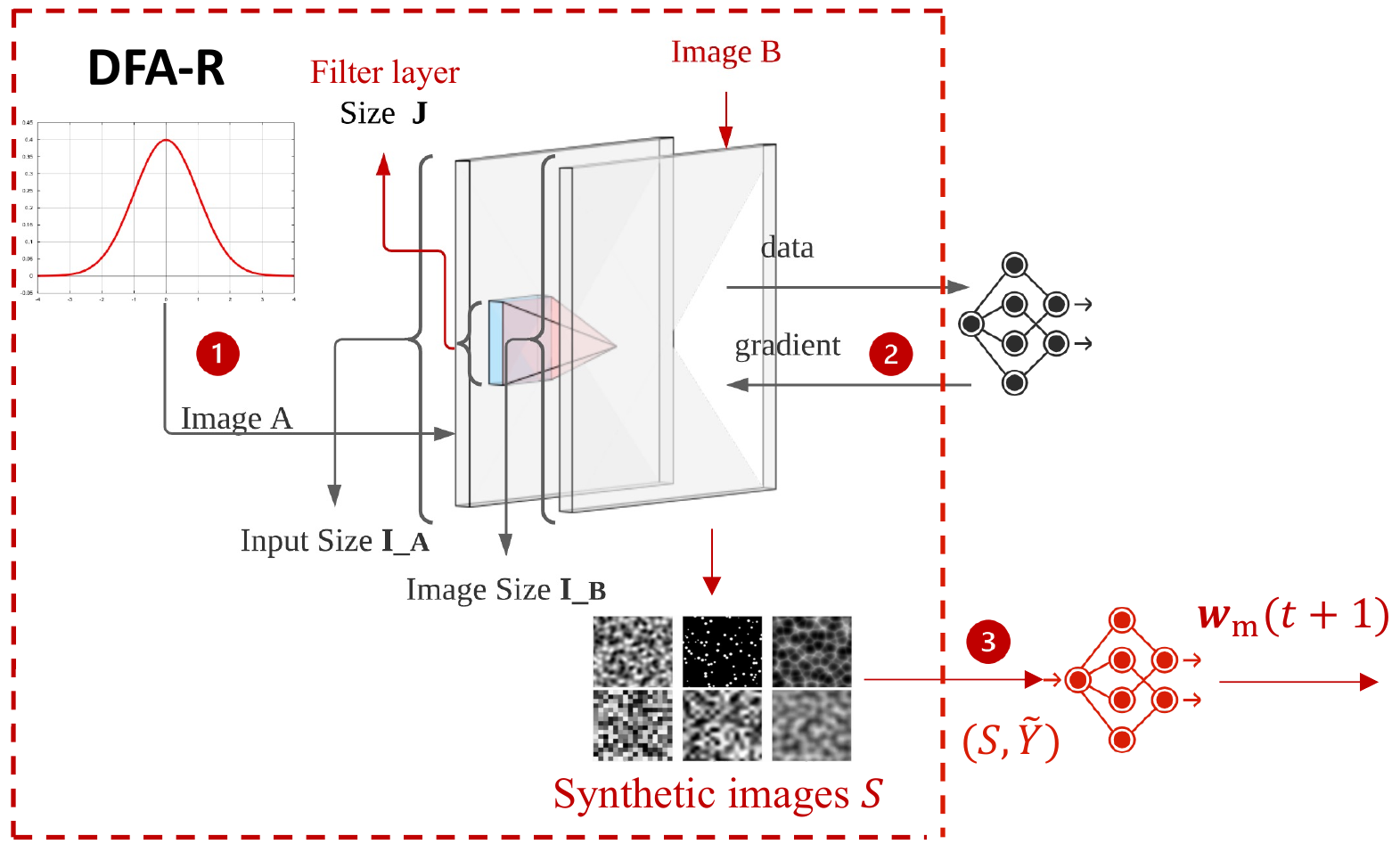}
\caption{Synthetic data generation process of \algl. }
\label{img:layerbased}
\end{figure}

When constructing the synthetic dataset $S$,  \algl aims to aggressively lower the confidence of all outputs of the global model by introducing a malicious filter layer(i.e., a convolutional layer). 
\algl optimizes this image layer such that per-class probability output of global model is equally low, i.e., $Y_D = [\frac{1}{L}, \frac{1}{L}, ..., \frac{1}{L}]$, where $L$ is the total number of classes. Such data is bound to confuse the global model. 
Fig.~\ref{img:layerbased} depicts the optimization procedure to find $|S|$ malicious images iteratively via two steps: \textit{i)} generating the malicious image through mapping a random dummy image via a filter layer~\cite{DBLP:journals/nature/LeCunBH15}, and \textit{ii)} optimizing the filter layer by minimizing cross-entropy loss of the global model
between the predicted class probabilities and $Y_D$. 

Concretely,  we first generate  a random image A (size $a\times a$), with each pixel being drawn from a uniform distribution,  and apply the filter layer to transform it into an image B(size $b\times b$).
In this manner, we train a mapping from randomness to images that have the desired properties. 
The size of image B is the same as the real image.
We let this convolutional layer have kernel size $J\times J$,
i.e., the square filter layer between image A and B in Fig.~\ref{img:layerbased}.  
 After being filtered from the convolution layer, the image B is then classified by the current global model. 
 \changesr{The attack works for various network structures and datasets, e.g., Alexnet, VGG on other image datasets, as long as the relation between input and output are maintained. Concretely, for stride size $St$ and padding size $P$~\cite{cnn:journals/tnn/LiLYPZ22}, we require $a= b \times (St+1) -2P+J$. }
\changesr{The attack can be extended to other tasks, e.g., text processing, by replacing the filter model and using a Seq2Seq model~\cite{sutskever2014sequence} instead of a convolutional layer. In this manner, the random text (mapped from random values to a dictionary) is filtered by the Seq2Seq model and fed in the text processing network, similar to Fig.~\ref{img:layerbased}.}
 


To optimize the convolutional layer that results in ambiguous $Y_D$, we first consider the dummy image A, the filter layer, synthetic image B, and the global model as one big classification problem.  Its training objective is to minimize the cross-entropy loss of predicted probabilities of image B and $Y_D$, such that the model cannot predict classes reliably. Different from the regular training of a classification problem, we keep certain parts of the model and input constant. Specifically, the model weights of the global model and the image A are static. Otherwise, without keeping A static,  we would need to re-train whenever we change the randomness. Keeping the number of trainable parameters to a minimum, we optimize the efficiency of the attack. 
The only trainable parameters here are the parameters of the filter layer. It takes $E$ epochs to train this convolutional layer. Upon finishing training, the image B is one data instance of $S$.  
To increase the diversity of the training dataset $S$, for each FL training round, we repeat the above process for $|S|$ 
times to construct $S$.

\subsection{\algg synthetic data generation}
\begin{figure}[thpb!]
\centering
\includegraphics[width=0.85\columnwidth]{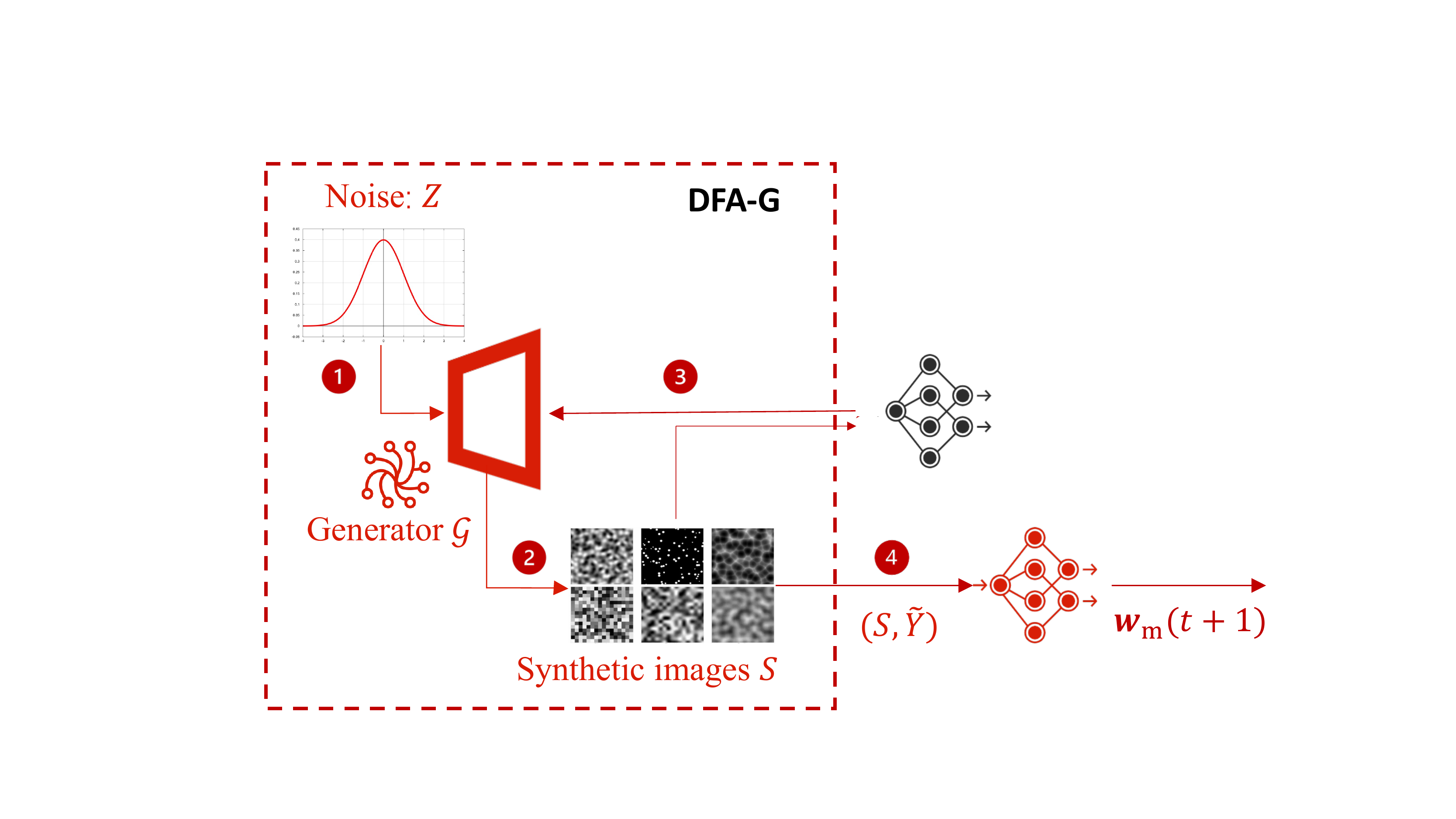}
\caption{Synthetic data generation process of \algg.}
\label{img:dfa_g}
\end{figure}

In contrast to \algl, \algg synthesizes images through a generator network, which misleads the global classifier to confidently make incorrect classifications.
In order to do so, we generate images that are not supposed to be from a class $\tilde{Y}$ but classify all of them as $\tilde{Y}$, which is a randomly chosen label and never changes through the training procedure.
The training/optimization of the generator network is through the feedback of the global model, i.e.,  we assume that the classification provided by the global model is the correct classification of the synthetic image.
Typically, benign training minimizes the cross-entropy of the prediction and true label so that the model can output an accurate prediction. However, as our goal is to reduce the model accuracy, we maximize the cross-entropy of the prediction and $\tilde{Y}$ to train $\mathcal{G}$, steering the generated images away from $\tilde{Y}$. \changesr{As \algl, \algg works for various network structures, with input and output size needing to match the training data.}

The training procedure for the generator is shown in Fig.~
\ref{img:dfa_g}. 
We first draw a random noise vector $Z$ from the Gaussian distribution and input $Z$ into
a generator $\mathcal{G}$ to synthesize malicious images. We use the same random seed over multiple rounds so that the trained generator is able to consistently produce
synthetic data different from class $\tilde{Y}$, \changesr{as our training goal is to optimize the mapping from the generated vector to the targeted synthetic data}. The network structure of the generator is a transpose convolution neural network (TCNN), which outputs task-specific image size data $S = \mathcal{G}(Z)$. The size of real data can be obtained from $\boldsymbol{w}(t)$. 
Specifically, we use a lightweight TCNN of two transposed convolutional layers and one convolutional layer following the structure of the popular WGAN paper~\cite{WGAN:journals/corr/ArjovskyCB17}. The model parameter of the generator $\mathcal{G}$ is randomly initialized before training, denoted as $\boldsymbol{\theta}$.
As the generator aims at synthesizing images  that differ from the 
chosen class $\tilde{Y}$, the objective function of $\mathcal{G}$ is $\max_{\boldsymbol{\theta}}F(\boldsymbol{w}(t), (S,\tilde{Y}))$ where $S$ is generated from $\boldsymbol{\theta}$.
After training $\mathcal{G}$ locally for $E$ epochs until convergence, the synthetic images are leveraged to train $\boldsymbol{w}_m(t+1)$. \changesr{When it comes to other tasks, e.g., text processing, the generator is a recurrent neural network such as GRU~\cite{GRU:journals/corr/ChungGCB14} in order to generate random texts rather than just random numbers.}



In summary, differences between the two variants are \textit{i)} the optimization network, \textit{ii)} the objective functions, \textit{iii)} the use of $\tilde{Y}$, and \textit{iv)} the randomness in their inputs.

\subsection{Distance-based Regularization}
\label{subsec:distance_loss}

State-of-the-art defenses in a FL system are mainly based on the pairwise distances among multiple updates. In order to bypass the defense mechanisms, we introduce a distance-based regularization term when training the adversarial classifier with the aim to further enhance stealthiness of the adversarial model updates. The concrete regularization term is
\begin{equation}
\label{eq:regularization}
  \mathcal{L}_d = \|\boldsymbol{w}-\boldsymbol{w}(t)\|_2 -  \|\boldsymbol{w}(t)-\boldsymbol{w}(t-1)\|_2. 
\end{equation}
In Eq.~\ref{eq:regularization}, the first term refers to the weight differences of the adversarial update and the current model. Analogously, the second term refers to the difference between the current model and the model of the previous round. \changesr{This term varies over rounds and the optimization variable is the model parameter $\boldsymbol{w}$. We add $\mathcal{L}_d$ to the vanilla cross-entropy loss in the objective function of the adversarial classifier to avoid extremely high differences in model changes over rounds, which could be easily detected.} Thus, in both \algl and \algg, we guide the training such that the differences in weights are similar to the ones in previous rounds, as achieved by using the two most recent global models. 

\section{Experimental Evaluation}
\label{sec:evaluation}



We empirically evaluate the effectiveness of our proposed data-free untargeted attack, on three commonly used classification benchmarks. We compare the attack success rate and defense pass rate for various settings,
for four state-of-the-art defenses and in comparison to three existing attacks. All of the results reported in this section are averaged over three runs. The source code of \alg is provided in github\footnote{https://github.com/GillHuang-Xtler/DSN2023DFA. For any question about code, please contact \{J.Huang-4, Z.Zhao-8\}@tudelft.nl.}

\subsection{Experiment Setup}

\textbf{FL system. }
Our FL system considered contains 100 clients.
In typical real-world FL systems, some of the clients could be offline or unavailable temporarily  and hence not all of them might be able to participate in the whole training process. 
Thus, as in previous work~\cite{backdoor:conf/aistats/BagdasaryanVHES20,fedavg:conf/aistats/McMahanMRHA17,ndss:conf/ndss/ShejwalkarH21}, 10 of the available clients are selected uniformly at random each round. Clients train the classifier locally for one epoch. For the main results, we assume that the adversary can compromise 20\% of the clients following~\cite{ndss:conf/ndss/ShejwalkarH21, Fang:conf/uss/FangCJG20}, unless stated otherwise, and further evaluate 10\% and 30\% in Sec.~\ref{subsec:proportion}. Lower percentages of attackers have been shown to be ineffective~\cite{shejwalkar2022back}.

\textbf{Datasets and networks.} In this work, we consider three datasets.
\textit{Fashion-MNIST}~\cite{xiao2017fashion} consists of a training set of 60,000 and a test set of 10,000 fashion-related images. Each instance is a 28 $\times$ 28 grayscale image. \textit{Cifar-10}~\cite{krizhevsky2009learning} contains 50,000 training images of 3-channel RGB images and 10,000 of test images. \textit{SVHN}~\cite{netzer2011reading} includes 73257 digit images for training and 26032 for testing. All digits have 32 $\times$ 32 pixels. \  All datasets have 10 classes in total. For Fashion-MNIST and Cifar-10, the images are evenly distributed over classes. SVHN in slightly imbalanced in class distribution. 
The total number of images used to train in this paper is reduced to 10\% for Fashion-MNIST and Cifar-10 but maintain the original size for SVHN. For Fashion-MNIST and Cifar-10, the data are chosen uniformly at random in order to model real-world scenarios that full data may not be available during the whole training. This amount is verified to be sufficient for training on Cifar-10 and Fashion-MNIST~\cite{subset:journals/pnas/BaldassiPZ20}.To determine the $|S|$ and show hyperparameter sensitivity, we run initial experiments varying $|S|$ from 20, 50, and 100 based on knowing 50 samples per client for Cifar-10. We found that \alg is able to achieve similar attack success rate. Indeed, sometimes a lower $|S|$ had a higher ASR, e.g., for DFA-G on Fashion-MNIST with $\beta=0.5$, $|S|=20$ has higher attack success rate than $|S|=50$. As they all succeed in attacking, we use the results of 50 in the paper to keep consistency. For these three datasets, we use representative neural networks with 2 (for Fashion-MNIST) and 6 (Cifar-10 and SVHN) convolutional layers connected with 1 and 2 densely-connected layers, respectively, to map the inputs and outputs\footnote{We use shallow networks to simplify evaluation, consistent with~\cite{ndss:conf/ndss/ShejwalkarH21}, higher accuracy can be achieved with deep nets.}.


\textbf{Defense mechanisms.} Four state-of-the-art defenses are evaluated in our work: \textit{mKrum}, \textit{TRmean}, \textit{Bulyan} and \textit{Median}. We do not apply \textit{Krum} since \textit{mKrum} interpolates between \textit{Krum} and averaging, thereby allowing the trade-off between the resilience properties and the convergence speed~\cite{krum:conf/nips/BlanchardMGS17}. 

\textbf{Data heterogeneity.}
To emulate
a heterogeneous distribution, we assign data to clients according to the commonly used Dirichlet distribution. It emulates a real-world data distribution and the degree of heterogeneity is governed by the  hyperparameter $\beta$~\cite{dirichlet:journals/bioinformatics/WuSWCC17},
indicating the level of heterogeneity. In Sec.~\ref{subsec:noniid}, we vary $\beta$ from 0.1 to 0.9 in order to demonstrate our effectiveness for different degrees of data heterogeneity. Higher $\beta$ means a lower degree of data heterogeneity. For our experiments, except for Sec.~\ref{subsec:noniid}, we choose $\beta = 0.5$, as in the prior work~\cite{dirichlet:journals/bioinformatics/WuSWCC17,dirichlet:journals/corr/abs-1909-06335}.

\textbf{Hardware.}
Our FL emulator is based on Pytorch and we run experiments on a machine running Ubuntu 20.04, with 32 GB memory, a GeForce RTX 2080 Ti GPU and an Intel i9 CPUs with 10 cores (2 threads each).

\subsection{Evaluation Metrics.}
We utilize two main metrics to evaluate the effectiveness of our attack.
\textit{i) Attack success rate (ASR)} is defined by:
\begin{equation}
    ASR = \frac{acc-acc_m}{acc}\times 100\%,
\end{equation}
i.e., the decrease of accuracy caused by attacks. Specifically, it is the difference between the global accuracy $acc$ without attacks and defenses and the maximum accuracy $acc_m$ of the global model during one experiment with attacks.
Attack success rate specifies the effectiveness of an attack strategy through the decrease in accuracy. The Higher, the better.

\textit{ii) Defense pass rate (DPR)} is a metric to measure the stealthiness of an attack. In our paper, it is defined by the proportion of attackers who have passed the defense ($N_p$) from all of the randomly selected attackers ($N_s$):\
\begin{equation}
    DPR = \frac{N_p}{N_s}\times 100\%.
\end{equation}
DPR as defined above requires that defenses select updates for aggregation rather than computes statistics on all updates. Thus, as detailed Sec.~\ref{sec:defensesSota}, DPR can only be computed for \textit{mKrum} and \textit{Bulyan}, but not for
\textit{TRmean} and \textit{Median}. High \dpr is better.

\begin{table*}[thpb!]
\renewcommand\arraystretch{1.2}
\centering
\caption{Attack success rate (\asr) and the maximum accuracy ($acc_m$) accordingly under attacks on Dirichlet distribution. $\beta = 0.5$. The accuracy without attacks and defenses $acc$ for Fashion-MNIST, Cifar-10 and SVHN is 82, 50, and 86, respectively, it is reasonable for our lightweight CNN~\cite{LIE:conf/nips/BaruchBG19}.}
\label{tab:asr}
\resizebox{1.75\columnwidth}{!}{%
\begin{tabular}{c|c|cc|cc|cc|cc|cc} 
\toprule
& & \multicolumn{2}{c|}{\textbf{Fang}} & \multicolumn{2}{c|}{\textbf{LIE}} & \multicolumn{2}{c|}{\textbf{Min-Max}} & \multicolumn{2}{c|}{\textbf{\algl}} &\multicolumn{2}{c}{\textbf{\algg}}\\
\textbf{Dataset} & \textbf{Defense} & acc (\%)& \asr(\%) & acc (\%)& \asr(\%) & acc (\%)& \asr(\%)  & acc (\%)& \asr(\%)  & acc (\%)& \asr(\%)\\ 
\midrule
Fashion-MNIST & \textit{mKrum} &73.5& \textbf{10.37} &         72.7      &           \textbf{11.34}&67.3  &\textbf{17.93}   & 52.6& \underline{\textbf{35.85}} & 64.3&   \textbf{21.59}     \\ 
& \textit{Bulyan} & 68.1&  \textbf{16.91}   &     75.0       &             {\textbf{8.54}}    & 56.8 &    \underline{\textbf{30.73}}  & 70.8  & \textbf{13.66}     & 59.8 & \textbf{27.07} \\ 
& \textit{TRmean}   & 30.9 & {\textbf{62.32}}  &      59.9          &             {\textbf{26.95}}       &37.8 &  \textbf{53.90}  & 21.9  & \underline{\textbf{73.29}}  &51.3 &  \textbf{37.44}    \\
& \textit{Median} & 61.1&   {\textbf{25.49}}&   73.4      &             {\textbf{10.49}}       &62.0 &   {\textbf{24.39}}  &62.0  & \textbf{24.39}   &60.9 & \underline{\textbf{25.73}}           \\ 

\bottomrule

Cifar-10 & \textit{mKrum} & 34.1& \textbf{31.80}    & 33.5    &    \textbf{33.00}      &    27.8 & \textbf{44.40}&    24.6 &  \textbf{50.80}  & 24.4 & \underline{\textbf{51.20}}  \\ 
& \textit{Bulyan}  &28.4 & \textbf{43.30}  &    31.4      &             {\textbf{37.20}}  & 21.2 &      \underline{\textbf{57.60}} & 22.2   &   {\textbf{55.60}}   &21.7 &{\textbf{56.60}}      \\
& \textit{TRmean}  & 13.9 &  \textbf{72.20}   & 13.1   &             {\textbf{73.80}}  &12.6 &  \textbf{74.80}     & 14.4  &   {\textbf{71.20}}   & 12.5 &\underline{\textbf{75.00}} \\ 
& \textit{Median}  & 24.5 &   {\textbf{51.00}}  &     37.0         &      {\textbf{26.00}}     &  24.9&  \textbf{50.20}      & 24.7  &\textbf{50.60}          &23.8  & \underline{\textbf{52.40}}   \\

\bottomrule

SVHN & \textit{mKrum} & 81.85& \textbf{4.83}    & 80.87    &    \textbf{5.97}      & 28.03& \textbf{67.41}&   60.33 &  \textbf{29.85} &26.36& \underline{\textbf{69.35}}  \\ 
& \textit{Bulyan}  &73.03 & \textbf{15.08}  &    50.18   &             {\textbf{41.65}}  & 19.66 &      {\textbf{77.14}} & 19.30   & \underline{\textbf{77.56}}    &42.70 &{\textbf{50.35}}      \\
& \textit{TRmean}  & 19.59 &  \underline{\textbf{77.22}}   & 46.76   &             {\textbf{45.63}}  &42.54 &  \textbf{50.53}     & 42.06 &   {\textbf{51.09}}   & 41.13 &{\textbf{52.17}} \\ 
& \textit{Median}  & 59.67 &   {\textbf{30.62}}  &     83.63        &      {\textbf{2.76}}     &  43.38&  \underline{\textbf{49.56}}      & 68.08  &\textbf{20.84}    &65.09  & {\textbf{24.31}}   \\
\bottomrule
\end{tabular}
}
\end{table*}

\textbf{Baselines. }We are the first to propose data-free untargeted attacks. So there is no direct baseline to compare to. To demonstrate the effectiveness, we compare our results with the three state-of-the-art attacks \textbf{LIE}~\cite{LIE:conf/nips/BaruchBG19}, \textbf{Fang}~\cite{Fang:conf/uss/FangCJG20} and \textbf{Min-Max}~\cite{ndss:conf/ndss/ShejwalkarH21} that require knowledge of benign updates or real data. We make the following choices regarding the parametrization of the defenses.
As defenses are unknown to the attacker in our scenario, we implement the version of the Min-Max attack that is designed for unknown defenses and achieves the best results.
For the Fang attack, the original paper assumed knowledge of the defense. 
We here use the version of the Fang attack that assumes
\textit{TRmean} or \textit{Median} as the defense, 
which is the only source code provided by the authors.
Otherwise, we use the parameters that produced the best results in the original papers.

\subsection{Comparison with baselines}
\label{sec:dfa_baseline}
\begin{figure}[ht]
\centering
\includegraphics[width=0.99\columnwidth]{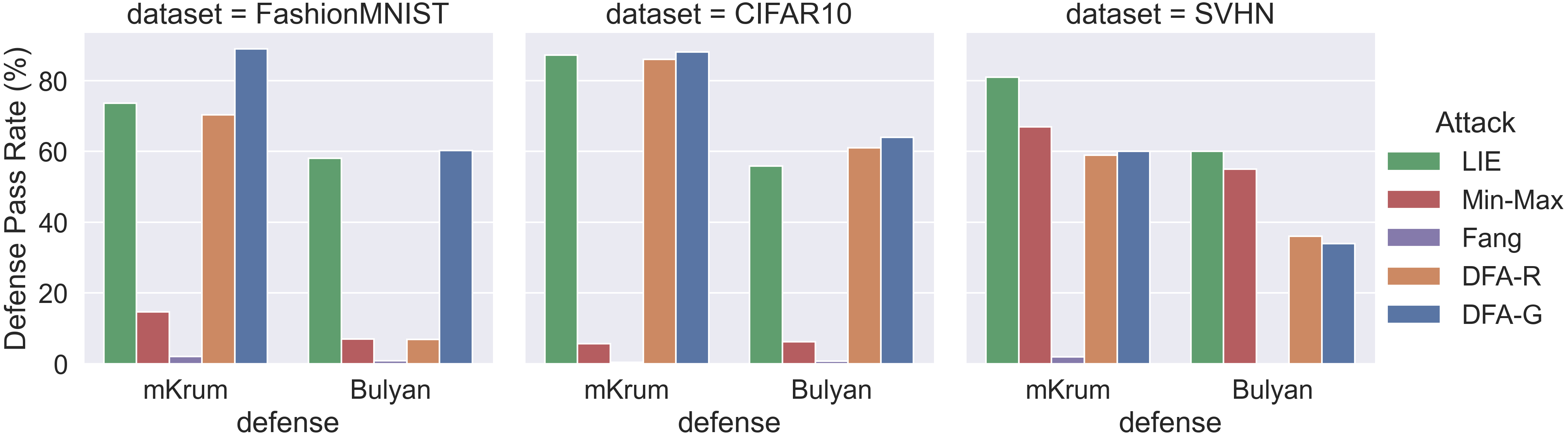}
\caption{Defense pass rate (\dpr) on Dirichlet distribution. $\beta = 0.5$ for Fashion-MNIST, Cifar-10 and SVHN. }
\label{img:dpr}
\end{figure}

\textbf{\asr and \dpr.} 
Our main results for the attack success rate and defense pass rate are shown in Tab.~\ref{tab:asr} and Fig.~\ref{img:dpr}. Among all of the baseline methods, Min-Max attack is the most successful attack, with high \asr even on low \dpr. In general, our experimental evaluation demonstrates that the proposed data-free attack strategies, \algl and \algg, are able to achieve similar or even slightly higher attack success rate than the baseline attacks, which require full knowledge of benign updates or a large quantity of raw data. \algg outperforms \algl in terms of \dpr for most results on different datasets, which  shows its stealthiness. During the first rounds of training, the attack is relatively weak as the global model does not yet provide a good enough model to generate effective poisoning, as indicated by our experimental results. Once the model converges, the polished model guides the attack and the attack success increases.

Specifically, from the results of Fashion-MNIST, \algl is better than \algg and all baselines when \textit{mKrum} and \textit{TRmean} are used to defend.  \textit{Bulyan} rejects on average more updates while \textit{Median} merely includes the median of each model parameter from all of the clients. Both make it hard to inject malicious data into the model, leading to the low pass rate for \algl and hence higher effectiveness of the more stealthy \algg. 
Correspondingly, \algl performs better 
\textit{mKrum} and \textit{TRmean} as they allow it to pass the defense more frequently.

On the other hand, \algg performs well for Cifar-10 due to the fact that training Cifar-10 networks with more layers (parameters) results in slower convergence so that it favours attacks that continuously circumvent the defenses. 
Also, the use of 3-channel RGB data increases the diversity of benign updates. 
As a consequence, the level of uncertainty is generally higher during training, so that it becomes easier to pass the defense as the benign updates are not consistent enough to act as a reference point that can be used to detect malicious images. 
For the same reason, \dpr of both \algg and \algl is higher on Cifar-10 than on Fashion-MNIST.
However, Fang and Min-Max are not more successful on Cifar-10. Min-Max, which is aware of benign updates and hence can adapt to different datasets, already integrates dataset-specific behavior that allows it to adapt to Fashion-MNIST's low diversity. Fang rarely passes defenses, regardless of the dataset. 
The results of \asr for Fang without knowing the exact defense is consistent with the original results~\cite{Fang:conf/uss/FangCJG20,ndss:conf/ndss/ShejwalkarH21}. 

For SVHN, both \algl and \algg achieve competitive \asr compared with the baselines. The only exception is Median where Min-max clearly outperforms our attacks. The result can be explained by the complexity of SVHN. SVHN is more complex than Fashion-MNIST, so it benefits from the additional knowledge Min-Max leverages to craft the update. 
In contrast, Cifar-10 also has a higher complexity, but experiments show that \textit{Median} has a low accuracy for Cifar-10 even in the absence of attacks if there is data heterogeneity, as it does not include important information in the model. So it does not make so much of a difference which attack is applied. The \dpr of \alg is lower than most other attacks for SVHN, in contrast to the other datasets. The results show that the effectiveness of the attacks depends on a combination of dataset and applied defense.   

We now consider the baseline attacks in more detail. LIE appears to be weaker than other attacks since it applies only a minor static shift to the mean of benign updates in order to pass defenses. This results in LIE's high \dpr but it limits its attack effectiveness. In contrast, Min-Max attack trains (maximizes) the scale of the shifting from the mean of benign updates each round so as to enhance effectiveness, especially under heterogeneous data. This the reason why it achieves good \asr even with low \dpr. The few times it overcomes the defense are sufficient for the crafted malicious updates to permanently damage the model. 
Fang attack has the least \dpr, as it steers the global model parameters to the reverse direction. It is even more easily detected by the defenses than Min-Max, to the extent that the attack effectiveness is severely reduced. 








\subsection{Data heterogeneity level}
\label{subsec:noniid}
\begin{figure}[h]
	\centering
	{
	\subfloat[Fashion-MNIST]{
	    \label{subfig:noniid_fashion}
	    \includegraphics[width=0.24\textwidth]{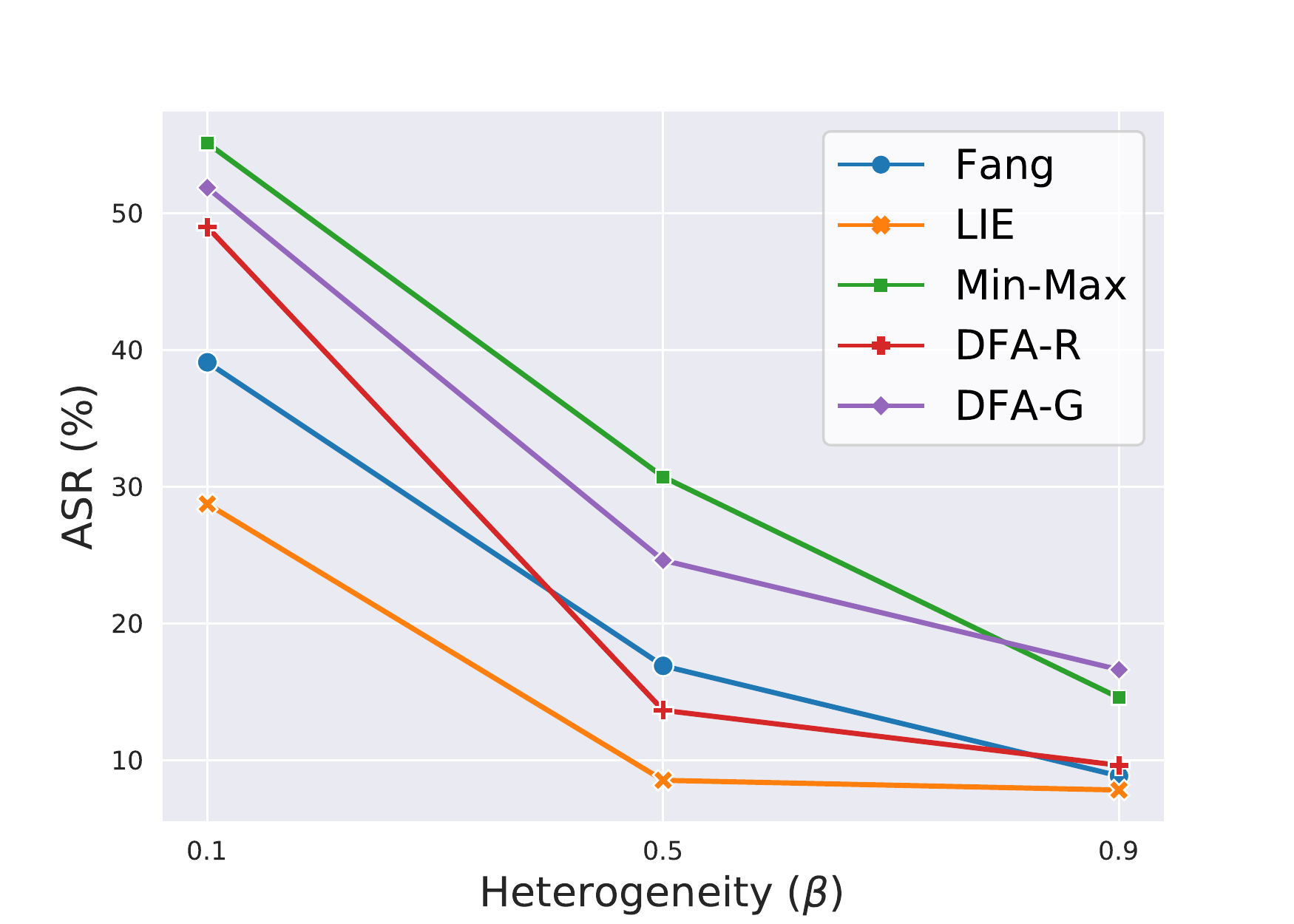} }
	    \hspace{-1em}
		\subfloat[Cifar-10]{
	    \label{subfig:noniid_cifar}
	    \includegraphics[width=0.24\textwidth]{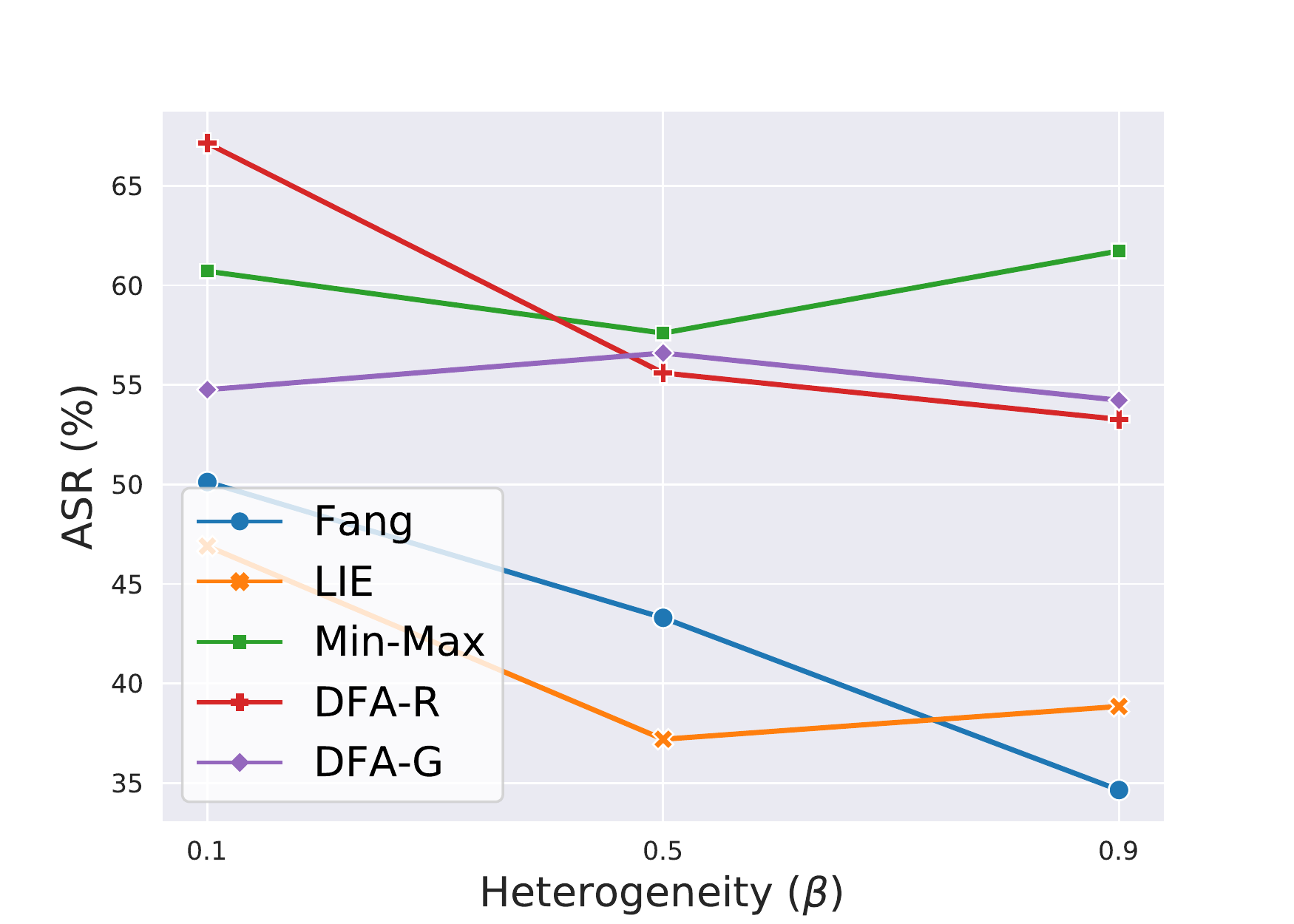}}
	}
	\caption{ASR(\%) for attacks under different levels of training data heterogeneity on Fashion-MNIST and Cifar-10 dataset.}
	\label{fig:noniid_level}
\end{figure}

We evaluate the impact of different levels of data heterogeneity on the \asr of attacks. Specifically, we choose $\beta = 0.1$ as the most heterogeneous case while $\beta = 0.9$ is the least heterogeneous case. 
Fig.~\ref{fig:noniid_level} displays the results for Fashion-MNIST and Cifar-10 when \textit{Bulyan} is used as a defense, which is a defense our attacks usually do not achieve the highest attack success rate,
as can be seen from Tab.~\ref{tab:asr}. In general, the effectiveness for all attacks increases with an increased level of data heterogeneity, since more heterogeneity means that the benign updates are more diverse and hence detection of outliers is harder.
The global model accuracy decreases on more heterogeneous data without attacks. This is consistent with the intuitive expectation that data of higher heterogeneity in an FL system results in poorer global accuracy within the same number of training rounds.

From Fig.~\ref{fig:noniid_level}, we can observe that for the aggressive \textit{Bulyan} defense, the Min-Max attack achieves mostly the best performance among all of the attacks. Attacks with full knowledge of benign updates as well as adaptive weights for maliciously shifting the mean is expected to work better. That is  especially true under aggressive defenses because in contrast to our attacks, Min-Max has access to information necessary to ensure their updates are less suspicious than others. 
Yet, thanks to the enhanced stealthiness, \algg outperforms Min-Max when data is less heterogeneously distributed among clients. Accordingly, \algl achieves the best results when $\beta=0.1$ on Cifar-10 dataset. In this scenario, the requirement of stealthiness is the least for all of the six scenarios because Cifar-10, as discussed above, has more diverse updates and the high degree of heterogeneity further increases the diversity, making it hard to detect outliers. 
Additionally, the \asr of LIE and Fang attack decreases drastically with decreased heterogeneity.  LIE attack adds a static minor shift to the true mean as it is designed to attack independent and identical distribution scenarios. For more heterogeneous updates, LIE attack is more likely to pass the defense and have an impact. 
Fang attack usually requires knowledge of the defense; in the absence of this knowledge, it fares better when its behavior is harder to be detected. \changesr{The results on SVHN dataset show similar trends with regard to data heterogeneity. As for Cifar-10, the \asr for $\beta=0.9$ may exceeds the \asr for $\beta=0.5$, e.g., \algg on SVHN has an \asr of 71.68\% and 50.35\% for $\beta=0.9$ and $\beta=0.5$, respectively.}






\subsection{Different proportion of attackers}
\label{subsec:proportion}
\begin{figure}[h]
	\centering
	{
	\subfloat[\textit{mKrum}]{
	    \label{subfig:fashion-maverick1-shapley}
	    \includegraphics[width=0.24\textwidth]{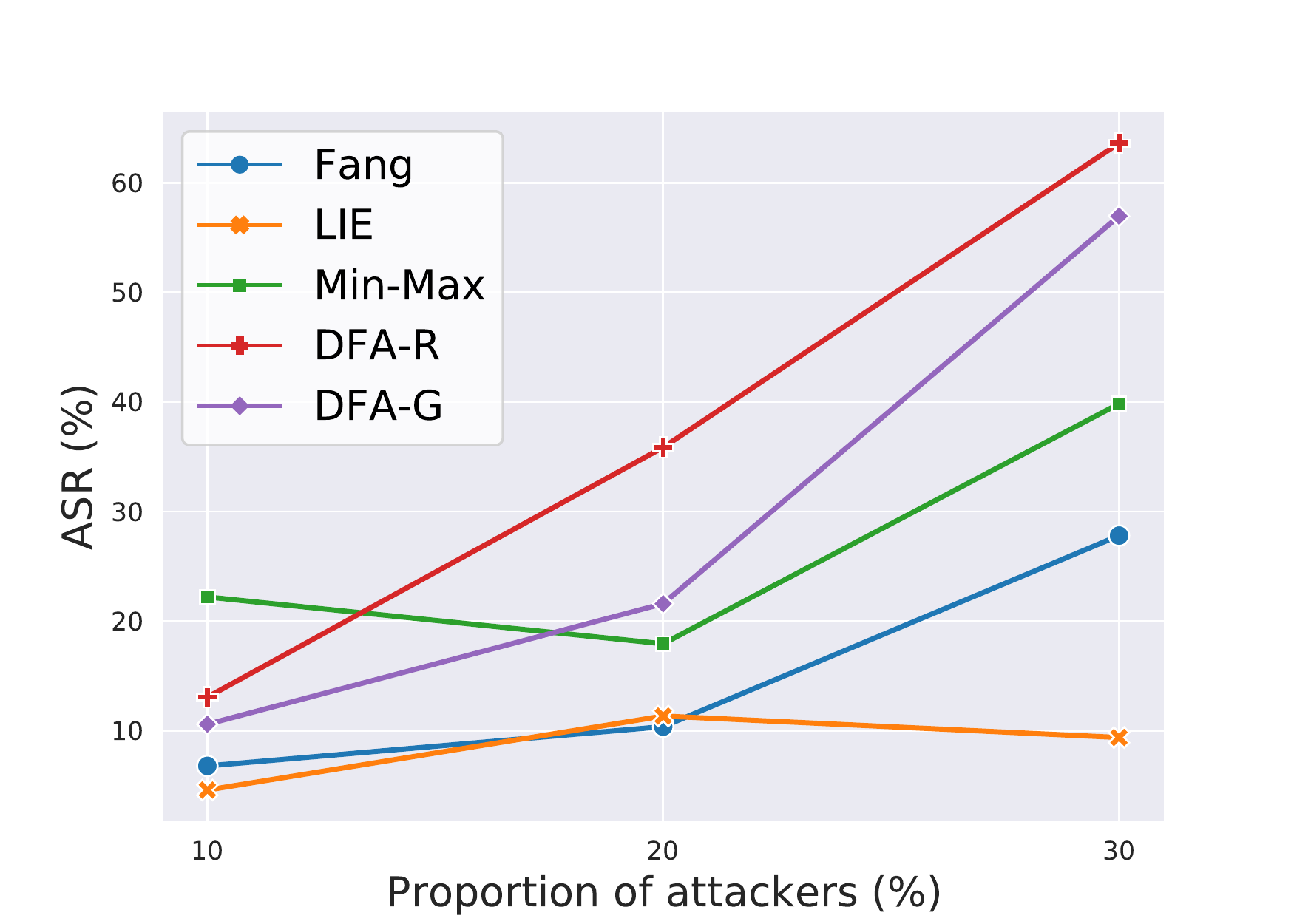} }
	    \hspace{-1em}
		\subfloat[\textit{TRmean}]{
	    \label{subfig:cifar-maverick1-shapley}
	    \includegraphics[width=0.24\textwidth]{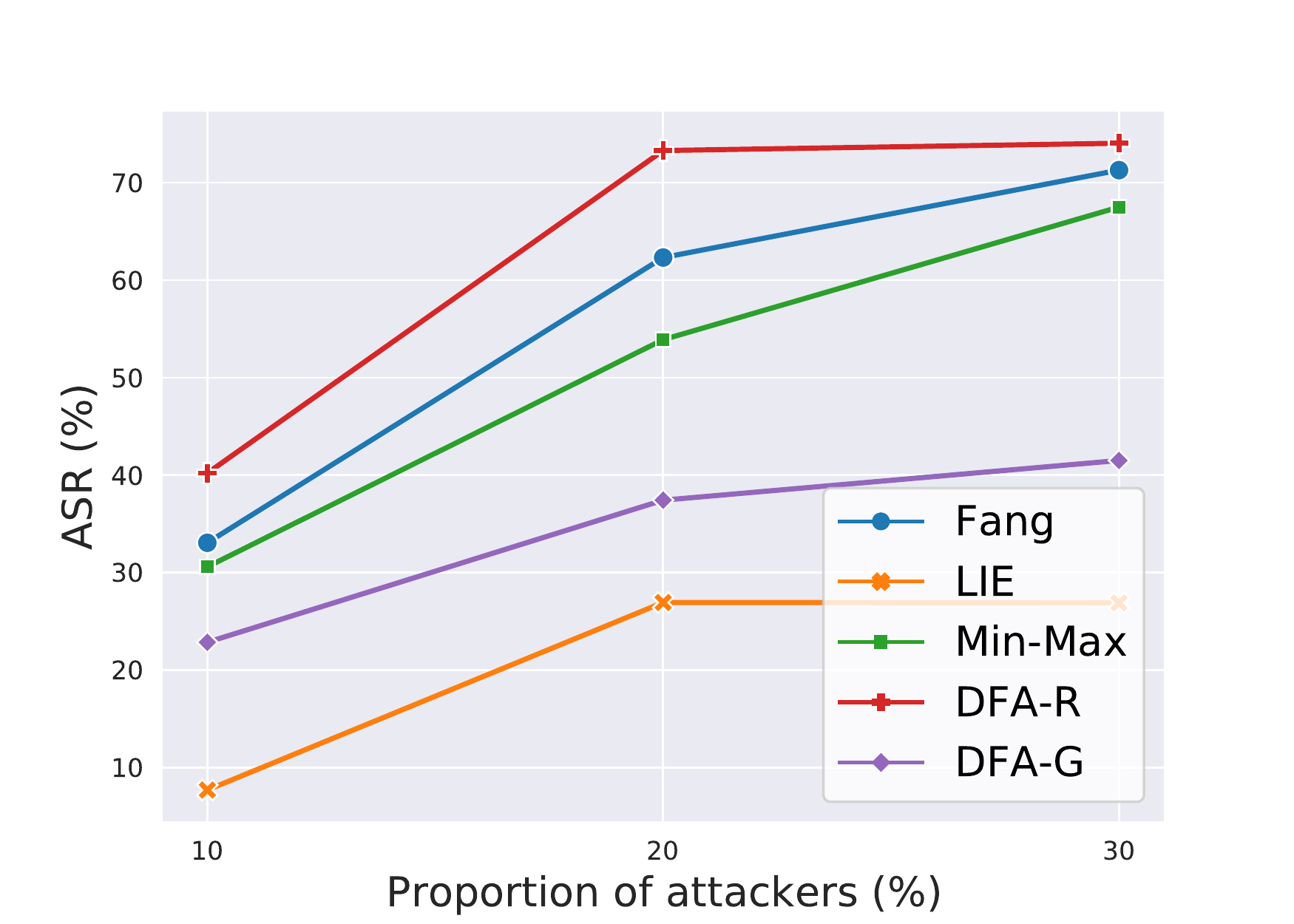}}
	}
	\caption{ASR(\%) for attacks under different proportions of attackers on \textit{mKrum} and \textit{TRmean} defense.}
	\label{fig:attack_proportion}
\end{figure}

In this section, we demonstrate the applicability of our proposed attack for different numbers of attackers. In order to show our effectiveness, we choose \textit{TRmean}, which is a statistic-based defense, and \textit{mKrum}, which is a distance-based defense, for our experimental results presented in Fig.~\ref{fig:attack_proportion}. The results are evaluated on the Fashion-MNIST dataset. We vary the attacker proportion from 10\% to 30\% as we do not expect the attackers of an FL learning system to exceed 30\%. To create heterogeneous data, we follow the  Dirichlet distribution with $\beta =0.5$ as in Tab.~\ref{tab:asr}. 






The results of Fig.~\ref{fig:attack_proportion}  demonstrate the consistent effectiveness of \alg compared with other attacks. 
The more attackers we have, the higher the attack success rate, as one expects. Yet, \alg achieves the highest attack success rate, compared to other attacks. 
\algl usually has the best performance, with the exception of 10\% on \textit{mKrum}, where Min-Max attack has the best \asr. 
Indeed,  it is easier to defend against a smaller number of attacks in the simpler dataset. As Min-max has more knowledge about the benign updates, it is then able to send in more malicious models than \alg in this easy-to-defend case. 


\subsection{Ablation analysis on \alg components} 





\subsubsection{Generator training epochs}

\begin{figure}[ht]
\centering
\includegraphics[width=0.99\columnwidth]{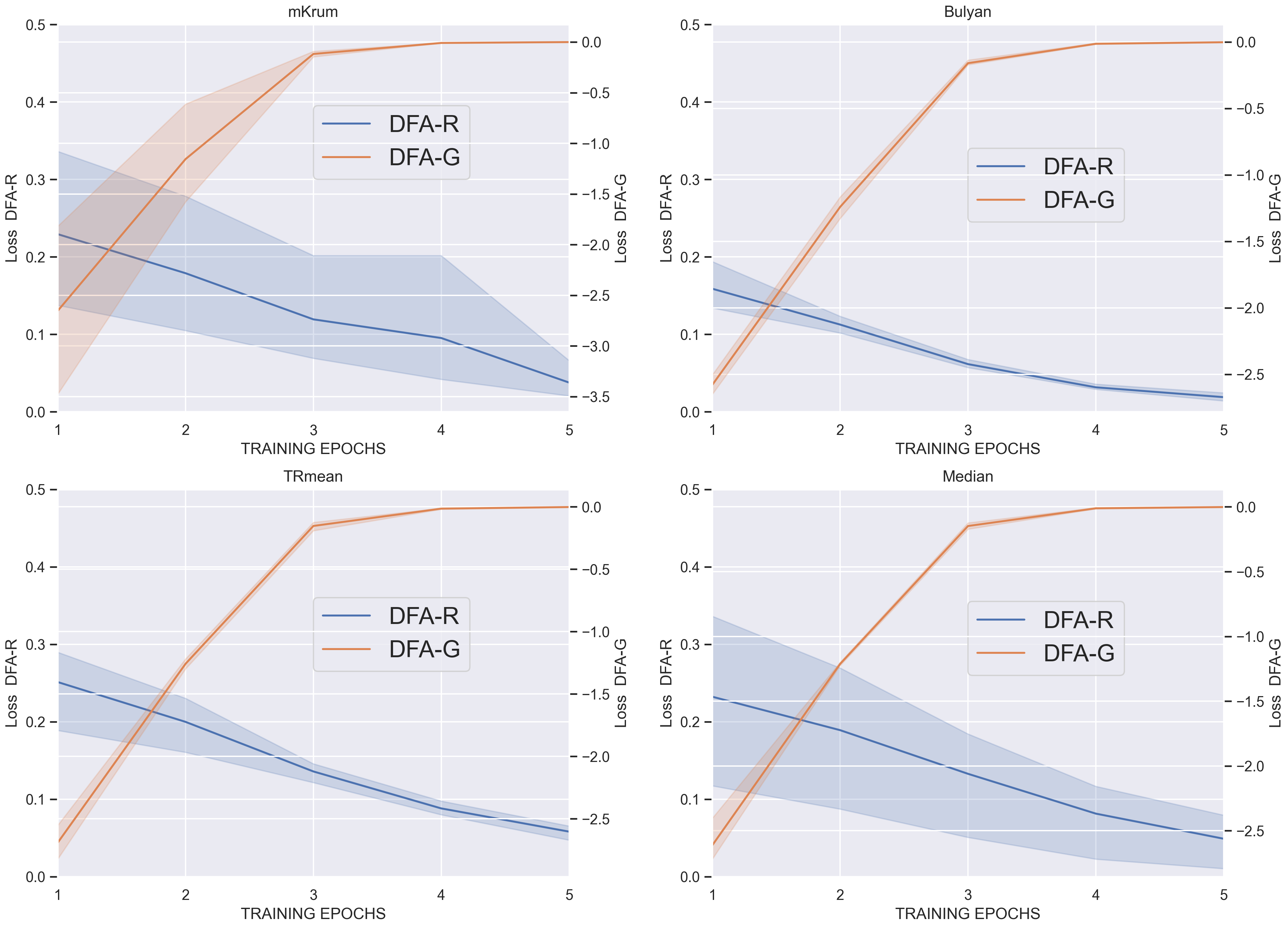}
\caption{Local training process of both \algg and \algl on Fashion-MNIST.}
\label{img:generator_train}
\end{figure}

Here, we empirically investigate the convergence towards the optimal loss, where \algl is minimizing its loss but \algg is maximizing it.
Fig.~\ref{img:generator_train} shows the results for Fashion-MNIST on all four defenses. 
It can be clearly seen that the local training for generating malicious images converges to a local optimum. For both of our proposed attacks, \algl and \algg, we only need a few epochs to train. For \algl, $E$ is 5 for Fashion-MNIST, and $E=10$ for Cifar-10 and SVHN as Fashion-MNIST is easier to train. 


\subsubsection{Comparison with non-training approach}
Given that the training converges fast, we also investigate the impact of training in comparison to just using a randomly initialized filter layer for \algl and a randomly initialized generator for \algg without any updating over rounds. As explained, according to the definition, \dpr is measured only on \textit{mKrum} and \textit{Bulyan} defenses. We hence report the results for \textit{TRmean} and \textit{Median} as ``N/A''. The maximum
accuracies without attack are the same as Tab.~\ref{tab:asr}.

The results can be seen in Tab.~\ref{tab:non-training} and confirm that training according to the current global model is indeed necessary. 
For \algl, training a single layer aims at generating images that confuse the global model.  Without the training step, the injection of \algl is less malicious.
Thus, \asr usually decreases without training, except for Fashion-MNIST with \textit{Bulyan} defense. This observation is due to the fact that  training \algl reduces the stealthiness of the attack by focusing on effectiveness and hence \algl passes \textit{Bulyan} more often without training. This is consistent with our results in Fig.~\ref{img:dpr} that \textit{Bulyan} significantly reduces the \dpr of \algl.

\begin{table}[htpb!]
\renewcommand\arraystretch{1.2}
\centering
\caption{\asr and \dpr for (non-)training approach where ``Static'' refers to non-training way with only randomly initialized. ``Fashion'' and ``Cifar'' is short for Fashion-MNIST and Cifar-10 datasets.}
\label{tab:non-training}
\resizebox{0.9\columnwidth}{!}{%
\begin{tabular}{c|c|cc|cc} 
\toprule
&  & \multicolumn{2}{c|}{\textbf{Static}} 
&\multicolumn{2}{c}{\textbf{Trained}}\\
\textbf{Attack} & \textbf{Defense} & ASR(\%) & DPR(\%) & ASR(\%) & DPR(\%)\\ 
\midrule

 & \textit{mKrum} & 18.17 &  87.78    &     35.85    &       70.33    \\ 
\algl& \textit{TRmean}   &37.20 &   N/A     &      73.29          &            N/A        \\
Fashion& \textit{Bulyan} & 23.66 &  57.50   &     13.66      & 6.86 \\   
& \textit{Median} & 21.22 &   N/A     &    24.39       &          N/A          \\ 
\bottomrule
 & \textit{mKrum} & 17.07 & 88.33 &       21.59     &   89.02    \\ 
\algg & \textit{TRmean}   &30.73&    N/A    &         37.44           &               N/A           \\
Fashion& \textit{Bulyan}  & 24.88 &  65.26&    27.07         &     69.33 \\      & \textit{Median} & 22.44 &     N/A     &    25.73        &             N/A         \\
\bottomrule

 & \textit{mKrum} & 50.00 & 85.20 &         50.80      &       86.04   \\ 
\algl& \textit{TRmean}   &71.14 &  N/A     &         71.20         &               N/A           \\
Cifar& \textit{Bulyan} & 56.00 &   60.98  &  55.65       &     61.05     \ \\    
& \textit{Median} &48.60 &    N/A      &   50.60       &       N/A         \\ 
\bottomrule
 & \textit{mKrum} & 38.60 & 56.46 &      51.20     &    88.14   \\ 
\algg & \textit{TRmean}   &71.40&    N/A    &        75.00          &       N/A                 \\
Cifar& \textit{Bulyan}  & 47.80 &  37.35&    56.60        &             63.99 \\      & \textit{Median} & 50.60&      N/A    &  52.40       &             N/A        \\
\bottomrule

\end{tabular}
}
\end{table}

When it comes to \algg, training helps to enhance stealthiness. The impact can be clearly seen from the results for \dpr in  Tab.~\ref{tab:non-training}, especially for \textit{Bulyan}. Only for a relatively lenient defense like \textit{mKrum}, the training has little additional impact as \dpr is already high without training.  
These results also reflect the minor increase of \dpr from Fashion-MNIST to Cifar-10 dataset for \textit{mKrum} in Fig.~\ref{img:dpr}.

\subsubsection{Impact of the regularization term}
In this part, we conduct an ablation study for our proposed distance-based loss, which adds a regularization term to the original cross-entropy loss function.
Tab.~\ref{tab:ablation_loss} shows both \asr and \dpr with and without the regularization term on Fashion-MNIST. For \algl, the effectiveness of the regularization term 
is more apparent for \textit{mKrum}. However, for \algg, the increase is most notable for \textit{Bulyan}. This is because the regularization term in the less stealthy \algl is insufficient for passing \textit{Bulyan} whereas it is what enables \algg to pass \textit{Bulyan} frequently. In contrast, \algg does not require the regularization term for \textit{mKrum} as it already passes without extra regularization.



\begin{table}[htpb!]
\renewcommand\arraystretch{1.2}
\centering
\caption{\asr and \dpr for ablation test of the regularization term proposed by our distance-based loss.}
\label{tab:ablation_loss}
\resizebox{0.9\columnwidth}{!}{%
\begin{tabular}{c|c|cc|cc} 
\toprule
&  & \multicolumn{2}{c|}{\textbf{without regularization}} 
&\multicolumn{2}{c}{\textbf{with regularization}}\\
\textbf{Attack} & \textbf{Defense} & ASR(\%) & DPR(\%) & ASR(\%) & DPR(\%)\\ 
\midrule

 \algl& \textit{mKrum} & 17.68 &  41.92   &     35.85    &     70.33  \\ 
& \textit{TRmean}   &58.78 &   N/A     &      73.29          &               N/A           \\
& \textit{Bulyan} &10.73 & 3.32  &     13.66      & 6.86 \\   
& \textit{Median} & 23.72 &   N/A      &    24.39      &          N/A         \\ 
\bottomrule
\algg& \textit{mKrum} & 20.98& 87.34 &       21.59     &   89.02    \\ 
 & \textit{TRmean}   &31.71 &    N/A    &         37.44           &               N/A          \\
& \textit{Bulyan}  &22.32&  60.27&    27.07         &     69.33 \\      & \textit{Median} &23.78 &     N/A     &    25.73        &             N/A        \\
\bottomrule
\end{tabular}
}
\end{table}

\subsubsection{Synthetic vs real data}
\begin{figure}[h]
\centering
\setlength{\abovecaptionskip}{-0cm}   
\includegraphics[width=0.35\textwidth]{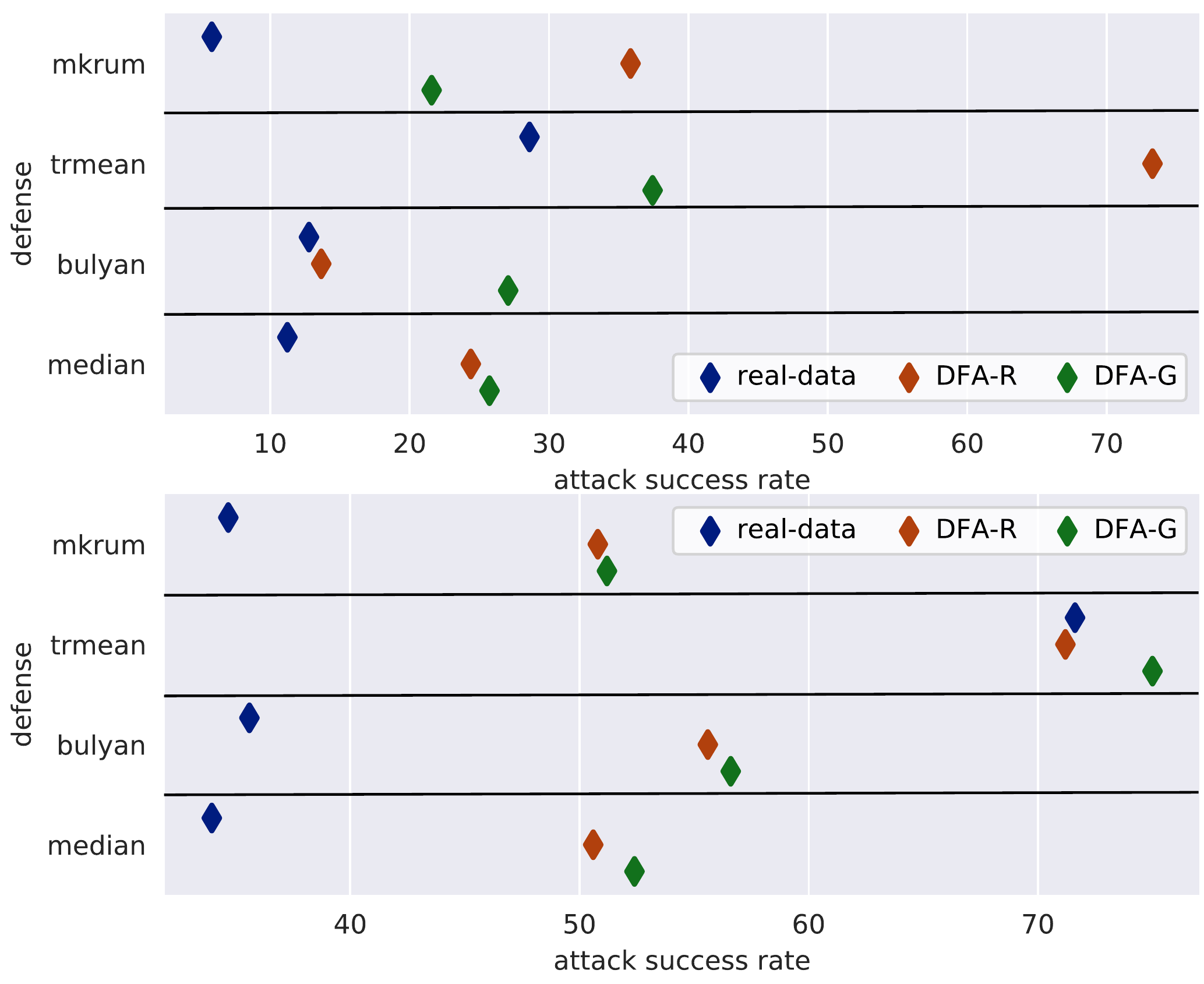}
\setlength{\abovecaptionskip}{0.1cm}
\caption{Comparison of \asr (\%) of real data and synthetic data by \algl and \algg with four defenses on Fashion-MNIST and Cifar-10.}
\label{img:comp_real_data}
\end{figure}

In order to demonstrate the effectiveness of our malicious synthetic data, we compare the \asr of our attacks to a version of the attack that uses real data, i.e., we use a set of real images instead of the synthetic image set $S$. 
We assign the number of real images owned by the attackers under the same Dirichlet distribution as for benign users. The results for the four defenses on both datasets are shown in Fig.~\ref{img:comp_real_data} with stripped visualization. ``Real-data'' in the figure refers to the results of \asr using real data paired with the uniformly chosen label $\tilde{Y}$ to train $\boldsymbol{w}(t)$
with distance-based loss as described in Sec.~\ref{sec:dfa} similarly for the synthetic data.
Fig.~\ref{img:comp_real_data} shows the effectiveness of our malicious synthetic data generated by \algl and \algg as \asr outperforms the case of using real images. That is expected because our synthetic images are specifically constructed such that the attack is very effective but at the same time stealthy. \changesr{Thus, even if data is present at the attacker, a data-free attack can be the better choice. Consequently, it is usually not necessary for the attacker to invest the overhead of obtaining data.}

\section{Defense for \alg: \algd}
\label{sec:defense}
Based on the results of the previous section, our attack is highly effective against known defenses. Yet, the attack might not withstand defenses that are crafted with data-free attacks in mind. Thus, in this section, we design and evaluate a novel defense that specifically addresses the reasons why existing defenses are insufficient. 

Let us first state why existing defenses fail. \textit{mKrum} and \textit{Bulyan} reject updates that differ greatly from others. Our regularization term ensures that our updates do not differ too much from the global model from the previous round, which at least once the model starts to converge is close to the models submitted by benign clients. Statistical methods lose information about the distribution but medians or trimmed means also shift easily without the need for an attacker to provide outlier data~\cite{LIE:conf/nips/BaruchBG19}. 

\subsection{Design of RefD (\algd)}
The defense is designed with data-free attacks in mind and overcomes the drawbacks of existing defenses. We rely on a reference dataset $\mathcal{D}_r$ to detect unusual classification patterns.  Upon receiving an updated model from clients, the server uses that model and  executes the model inference on $\mathcal{D}_r$, i.e., they compute predicted class probabilities. We design a novel statistic, $D$-score, which identifies the model updates whose outputs have either biased prediction or low confidence. To evaluate the effectiveness of \algd, we apply \algd on both our proposed attacks and the state-of-the-art attacks, for both balanced and heterogeneous data distributions.

\textbf{Assumptions. }\algd is a server-side defense. \algd requires that the server owns a small reference set $\mathcal{D}_r$ of real data with correct labels. The quantity of each class label is assumed to be balanced.

The design core of \algd is the $D$-score for each model update received: a low score indicates a high risk of update being malicious and results in the server  rejecting the update. The $D$-score is computed based on two parts: the balance value and the confidence value of updates. The balance value determines how balanced the outputs from the updated model are, to detect updates that are biased toward a specific class, such as  \algg, LIE~\cite{LIE:conf/nips/BaruchBG19}, and Min-Max~\cite{ndss:conf/ndss/ShejwalkarH21}.
The confidence value measures the confidence in predicting a class and rejects updates that result in low confidence, which are the objective of \algl and Fang~\cite{Fang:conf/uss/FangCJG20}.  


Before explaining those values, we first explain background notations.  $\boldsymbol{w}_i$ defined as the updated classifier model received from the client $i$,  which maps the data input into two kinds of output. Specifically, $\boldsymbol{w}_i^p(\cdot)$ maps the data into the per class probability vector and $\boldsymbol{w}_i^h (\cdot)$ maps the data into the per class one-hot encoding vector. Both vectors have the length of $L$, corresponding to the number of classes.

\textbf{Balance value.} To tackle the attack type which causes bias in classification, we define the balance value $B_i$ for the updated model of client $i$ as the inverse of the standard deviation of the class label distribution: 
\begin{equation}
    B_i = \left\{
    \begin{aligned}
    \frac{1}{std(A_i)}, \textnormal{if }  std(A_i) \neq 0\\
    1, \textnormal{if } std(A_i) = 0
    \end{aligned}
    \right.
\end{equation}
where $std(\cdot)$ is the standard deviation over all class labels and  $A_i$ consists of the aggregated number of predicted labels of each class. For instance, $A_i$ for Cifar-10, is a set of 10 values, i.e., the number of predicted samples per class. To compute $A_i$, we first apply $\boldsymbol{w}_i^h$ on each sample in $D_r$ and aggregate them.
Overall, the non-biased output prediction from benign clients results into a better balanced $A_i$ across all classes and thus a higher value of $B_i$. The adversarial parties on the other hand should have a lower value of $B_i$. 

\textbf{Confidence value. }This value quantifies the average confidence when using the updated model on $\mathcal{D}_r$. 
Specifically, for a data sample $j$ in $\mathcal{D}_r$, we let the confidence of applying classifier of client $i$, $M_{ij}$ as the biggest element of the output probability vector $\boldsymbol{w}_i^p(\mathcal{D}_r(j))$. We thus define the 
confidence value, $V_i$, as:
\begin{equation}
    V_i = \frac{1}{|\mathcal{D}_r|}\sum_{j=1}^{|\mathcal{D}_r|} M_{ij}. 
\end{equation}
Low confidence values indicate higher risks of being adversarial updates. The potential drawback of using $V_i$ to detect adversarial behaviour is that the low confidence may also appear in the early training epochs of honest clients.

\textbf{D-Score. }
For each client  update, we  combine the balance value, $B$, and the confidence value, $V$, to detect a wide range of adversarial behaviors. 
Motivated by $F_\beta$ Score~\cite{sasaki2007truth}, we define the $D$-Score to evaluate the quality of an intermediate training model from client $i$ as:
\begin{equation}
    D-Score = (1 + \alpha ^2) \times \frac{B_i \times V_i}{\alpha ^2 B_i + V_i},
\end{equation}
where $\alpha$ is a hyper-parameter to weigh the importance between balance value and confidence value. It can be set as a specific value according to what the central server knows or suspects about the executed attack. It can also be adaptive and learned over epochs, but we consider this out-of-scope for the paper and a good avenue for future work. Instead, we set $\alpha=1$ to represent the equal importance of $B_i$ and $V_i$.
If the predictions are perfectly balanced and have a high confidence, we have a D-Score of 1. When $B_i$ is reduced while $V_i$ stays constant, the D-Score is reduced, mirroring the increased bias. Analogously for $V_i$, a lower value for $V_i$ leads to a lower $D$-Score to indicate the lower confidence.  Moreover, as we designed the defense with data-free attacks in mind,  we expect it to work better for those than for other attacks, for which defenses already exist.

\textbf{Removing attackers. }After calculating the $D$-Score for each update of a given round, the server rejects the
updates with the $X$ lowest $D$-Scores. 
The server then excludes them for aggregation. 
The method is the same as used by \textit{mKrum}~\cite{krum:conf/nips/BlanchardMGS17}). $X$ is determined by the server's assumptions about the fraction of attacker, i.e., the more attacker they expect, the higher they choose $X$. 

\subsection{Evaluation for \algd}

\textbf{Experimental settings. }
To evaluate the effectiveness of \algd, we compare the accuracy of the global model in the presence of the new defense. We use the full test set 
for the respective dataset in the presented results but also experimented with smaller reference datasets (1000 images instead of 10000) and found no significant difference. Hence, smaller datasets can be used to increase efficiency and lower the requirements in terms of data availability at the server side. 
However, the reference set has to be balanced among class labels to compute the balance value reliably. \algd is evaluated on both Fashion-MNIST and Cifar-10 dataset. Additionally, our experiments include four different levels of data heterogeneity: independent and identical distributed (\textit{i.i.d}) and three heterogeneity levels ($\beta = 0.1, 0.5, 0.9$, where $\beta = 0.1$ indicates the highest level of heterogeneity) as in Sec.~\ref{subsec:noniid}. 
We also evaluate the impact of different level of data heterogeneity since defenses are sensitive to the heterogeneity, especially for distance-based defenses. Intuitively, a higher level of training data heterogeneity makes defense more difficult. The robustness of a defense in presence of high data heterogeneity is important to various of real-world application scenarios.

As in other works~\cite{Fang:conf/uss/FangCJG20, ndss:conf/ndss/ShejwalkarH21}, we set the proportion of attackers in the system to be 20\% and $X = 2$.
We compare \algd against \textit{Bulyan}, the most effective SOTA defense for our attack.






\textbf{Results for defense \alg. }

\begin{figure}[t]
	\centering
	{
	\subfloat[Fashion-MNIST]{
	    \label{subfig:fashion-result}
	    \includegraphics[width=0.24\textwidth]{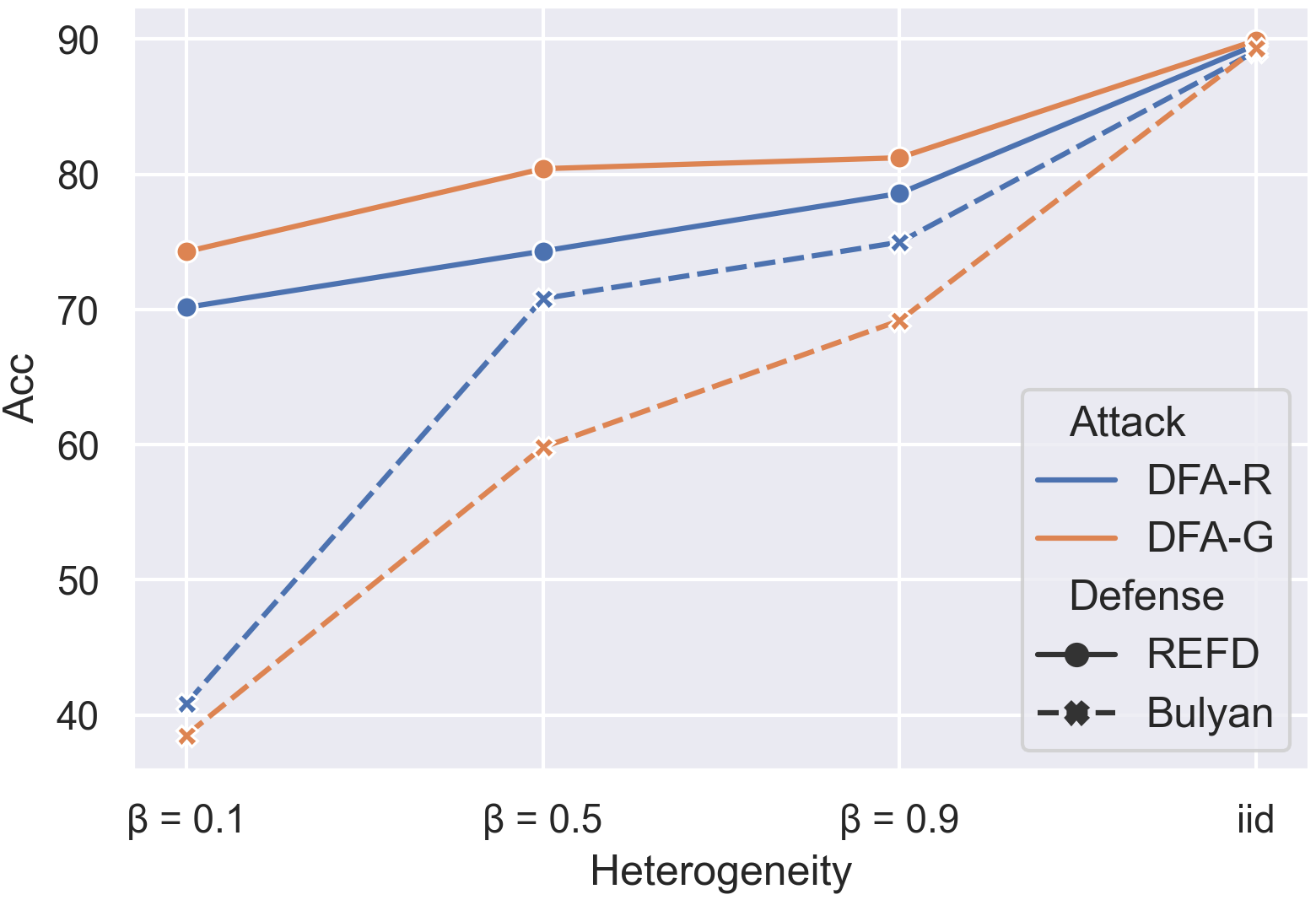} }
	    \hspace{-1em}
		\subfloat[Cifar-10]{
	    \label{subfig:cifar-result}
	    \includegraphics[width=0.24\textwidth]{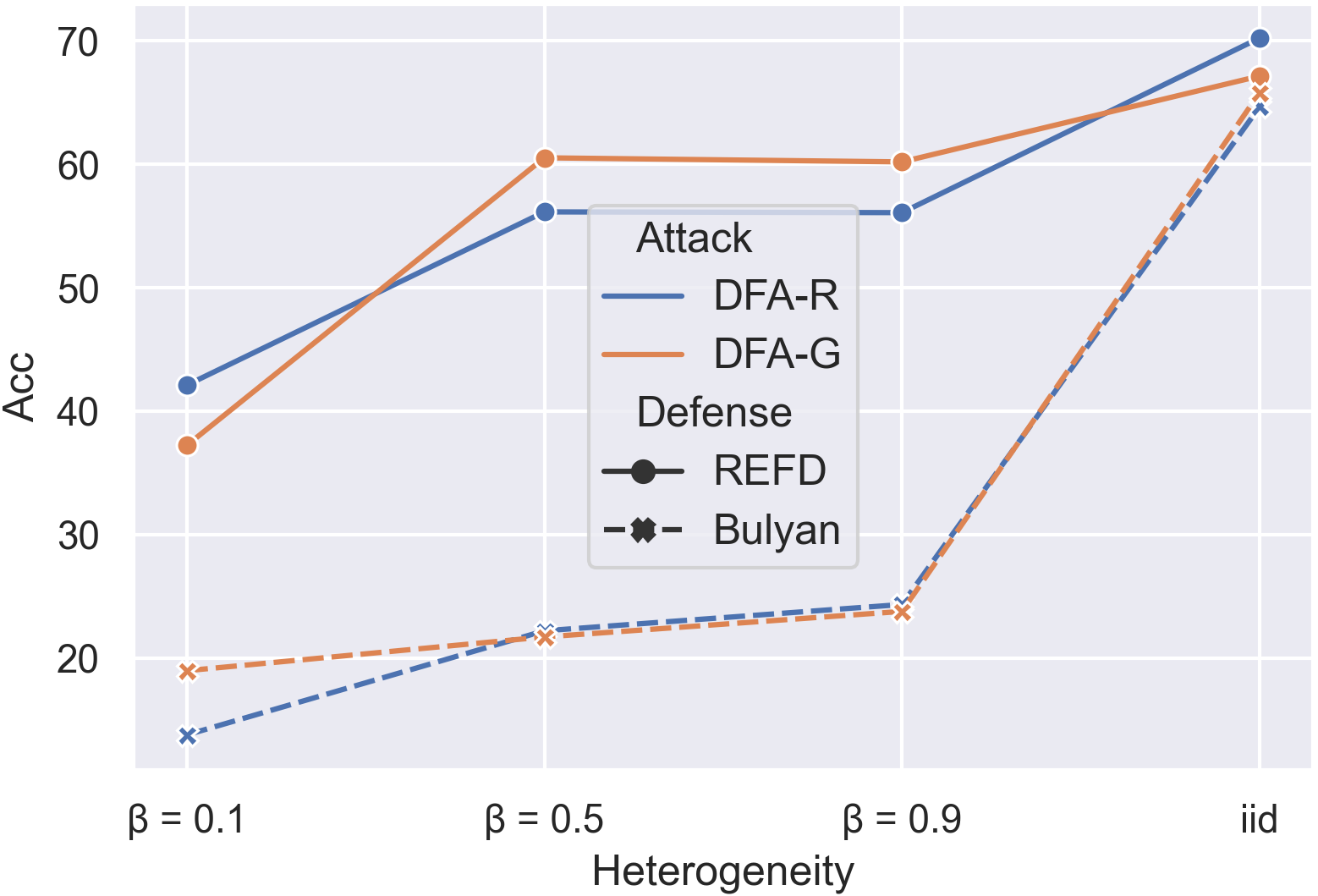}}
	}
	\caption{Accuracy(\%) for \algd on Fashion-MNIST and Cifar-10 datasets with different levels of data heterogeneity, compared with the maximum accuracy under \textit{Bulyan} defense.}
	\label{fig:dfad_heterogeneity}
\end{figure}

\begin{figure}[h!]
	\centering
	{
	\subfloat[Fashion-MNIST]{
	    \label{subfig:fashion-result-7}
	    \includegraphics[width=0.48\textwidth]{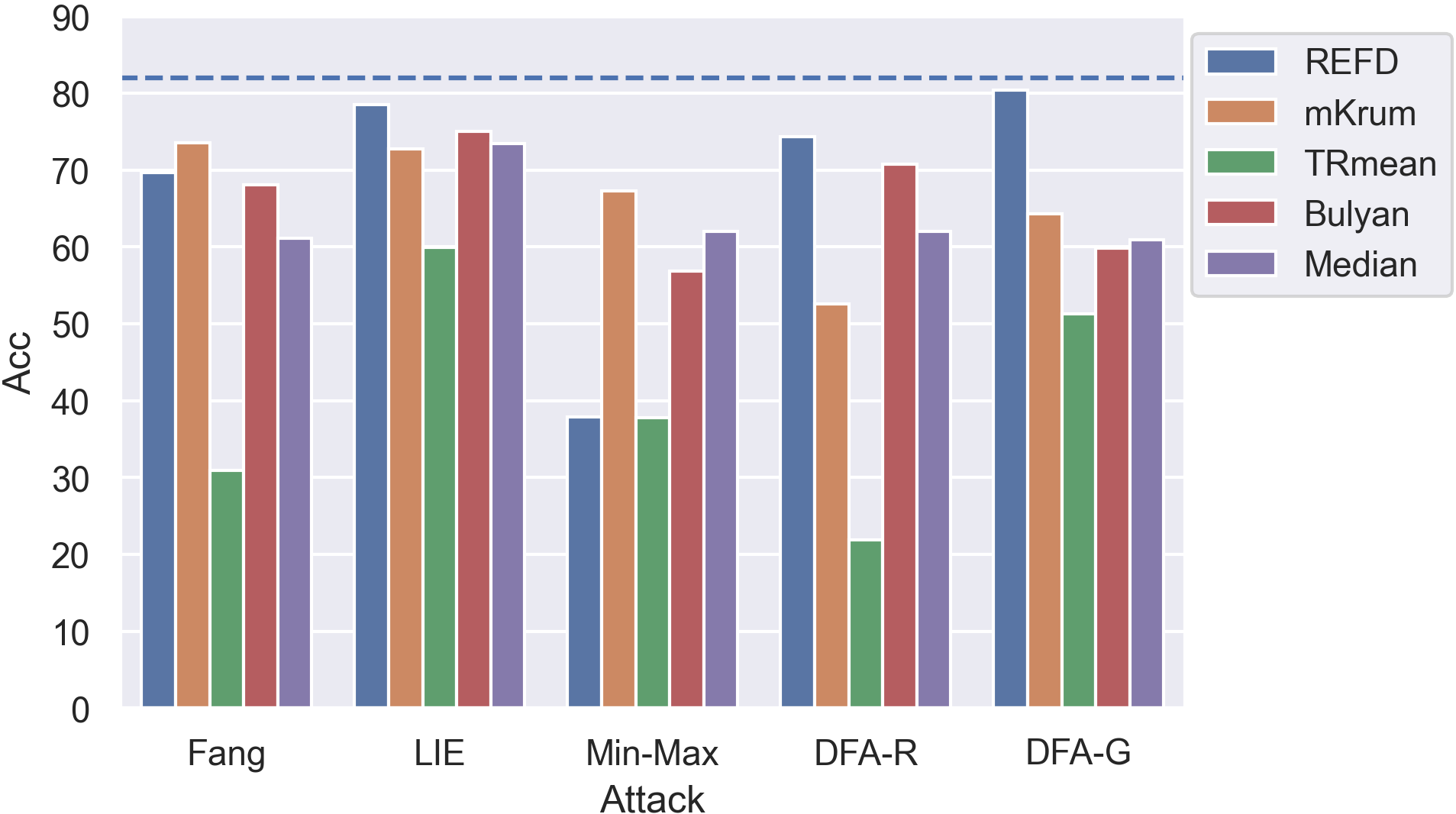} }
	    \hspace{-1em}
		\subfloat[Cifar-10]{
	    \label{subfig:cifar-result-7}
	    \includegraphics[width=0.48\textwidth]{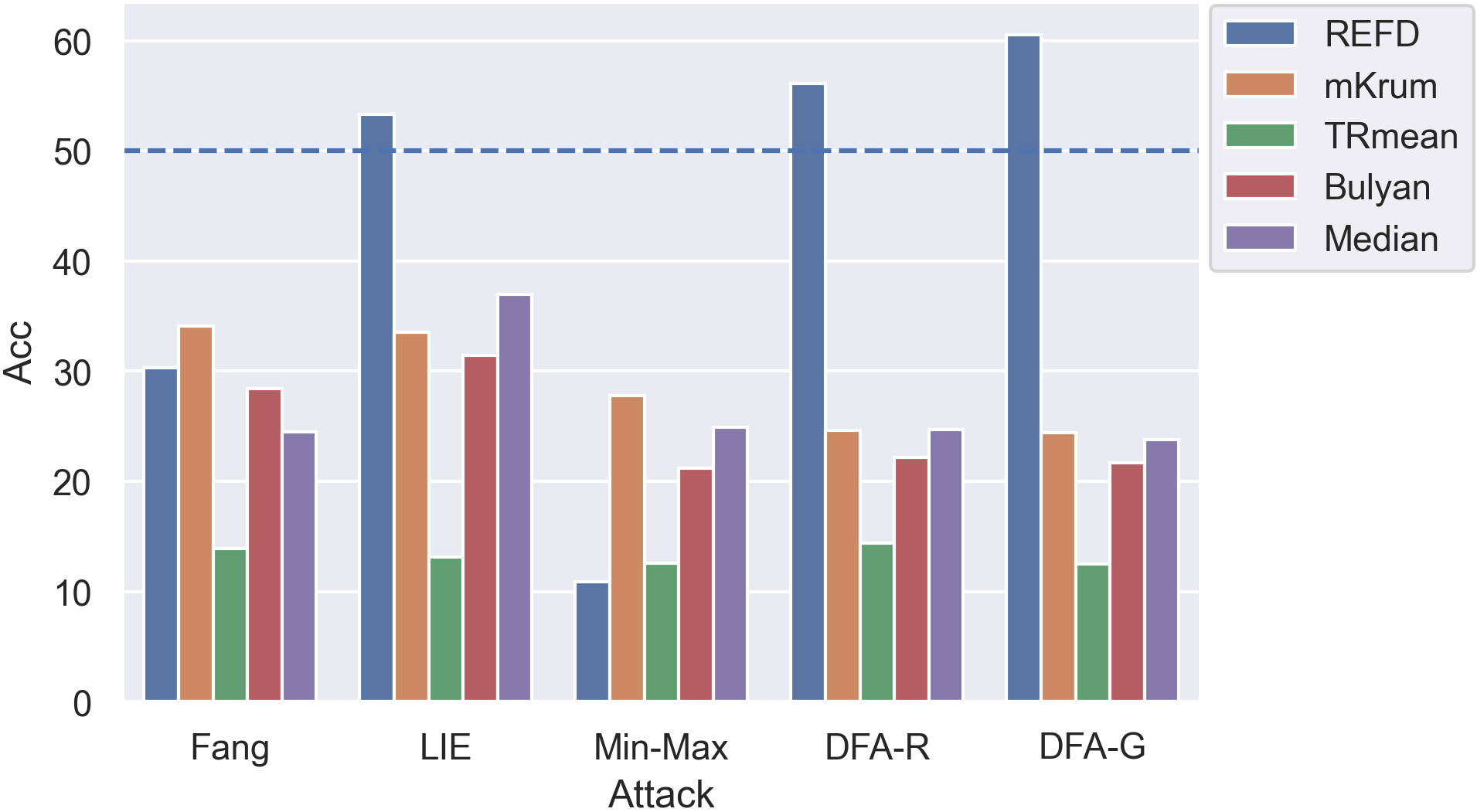}}
	}
	\caption{Comparison of Accuracy(\%) for defenses with state-of-the-art attacks on Fashion-MNIST and Cifar-10 datasets. }
	\label{fig:dfad_comp}
\end{figure}

From Fig.~\ref{fig:dfad_heterogeneity},  we see that \algd  significantly outperforms \textit{Bulyan}. 
The advantage of our defense is obvious when the heterogeneity of the data is high.
For $\beta=0.1$ and Fashion-MNIST, \algd achieves an accuracy of more than 70\% when the attack is \algl and close to 75\% for \algg. In contrast, the accuracy of \textit{Bulyan} is only around 40\%.
\textit{Bulyan} is relatively effective for \textit{i.i.d} data, achieving a similar value as \algd for both attacks.






The results on Cifar-10 confirm the superiority of \algd. 
As noted in Sec.~\ref{sec:evaluation}, the accuracy on Cifar-10 is generally lower. Yet, \algd is clearly the more effective defense. In the presence of data heterogeneity, the accuracy is at least twice as high for \algd than \textit{Bulyan}.

For both datasets, the advantage of \algd is least pronounced for the \textit{i.i.d} setting.
For \textit{i.i.d}, the benign updates are very similar, hence making it barely possible for our attacks to deviate without being detected. Thus, there is little difference between defenses.  
The results for Fashion-MNIST and Cifar-10 differ in that the final global model accuracy difference between \algd and \textit{Bulyan} is larger for Cifar-10. The data complexity of Cifar-10 is higher, increasing the difficulty of defending, so that our specifically designed defense shows a more pronounced advantage. 

The achieved accuracy is close to the accuracy achieved without attacks and defenses. For instance, for $\beta=0.5$, the accuracy without attacks and defenses is 82\% for Fashion-MNIST, which is less than 2\% higher than the accuracy for \algd for \algg. For \algl, the attack decreases the accuracy by about 10\%. 
For Cifar-10, the accuracy with and without attacks are almost equal. Indeed, the accuracy with attack and defense is insignificantly higher for some settings, which is likely due to randomness.

\textbf{Defending against other attacks. }
\algd is designed with \alg in mind, as our goal is to show that we can defend against data-free attacks. The design does not necessarily work against all attacks. Here, we establish whether the defense can nevertheless defend against Fang, LIE, and Min-Max attack \changesr{ and compare it with the results for \alg.}

The results are reported in Fig.~\ref{fig:dfad_comp}. We also follow the setting of 20\% attacking proportion and compare the maximum global model accuracy for the state-of-the-art defences and \algd. \changesr{We include the baseline accuracy with no attack and no defense as the dashed line.} From Fig.~\ref{fig:dfad_comp}, we can see that \algd has a good defending performance in general. However, it is not always the best among the-state-of-the-art defenses. Specifically, for LIE attack, \algd gets the best defending performance. LIE shifts the true statistical features of the benign updates, which can easily be caught by the balance value $B$ of \algd. 
Moreover, \algd also protects the model from Fang attack, where it achieves the second best ranking on both datasets. 
\algd works well for Fang since Fang updates malicious models on the opposite direction, which causes low confidence, i.e., low $V$. 
However, \algd is less effective against Min-Max than other defenses, as Min-Max's scaling technique should not affect balance and confidence value much. In summary, \algd protects well against data-free attacks presented in this paper and can also protect against other attacks. However, it is not a generic defense and hence should be applied in combination with other defence mechanisms. \changesr{It is also interesting to note that with RefD, the global model accuracy can even be higher than the baseline on Cifar-10. This result implies that RefD has benefits in the presence of data heterogeneity in comparison to FedAVG.}

\subsection{Overhead analysis for defense}
The defense does not add any communication, so merely the computation complexity is affected. The defense first evaluates the local update of each client for each image in the reference dataset, so the cost is  $\mathcal{O}(|\mathcal{D}_r|K)$ times the cost of evaluating the update. 
Furthermore, the $D$-Score needs to be computed, which is linear in  $\mathcal{O}(|\mathcal{D}_r|)$ as we compute the standard derivation (for $B$) and the maximum (for $V$) of $\mathcal{O}(|\mathcal{D}_r|)$ values. Last, we determine the clients with the smallest values, which has complexity $\mathcal{O}(K)$. 
Overall, evaluating updates is of a lower complexity than training new models, so the overhead is not prohibitive and can be reduced by using a smaller set $\mathcal{D}_r$. 





\section{Conclusion}

We propose \alg,  the first data-free untargeted attack on FL. Our results confirm that data-free attacks can be similarly or even more effective than other attacks that require data or benign updates, due to generating synthetic images to train on that are particularly useful at steering the model into the wrong directions. Furthermore, we design a defense strategy \algd that effectively protects against the proposed \alg and existing attacks by leveraging the statistics of model outputs in predicting reference data. In the future, we want to \changesr{explore \alg on different data types, e.g., text}, and check whether combining synthetic and real data in an attack can improve attack effectiveness and to what extent data is needed in a defense. 

\footnotesize
\bibliographystyle{plain}
\bibliography{main}

\end{document}